\documentclass{aa} 

\usepackage{graphicx}
\usepackage{txfonts}
\usepackage{soul}
\usepackage{xcolor}

\begin{document}

  \title{Clean catalogues of blue horizontal-branch stars using Gaia EDR3\thanks{The catalogues are only available in electronic form at the CDS via anonymous ftp to cdsarc.u-strasbg.fr (130.79.128.5) or via http://cdsweb.u-strasbg.fr/cgi-bin/qcat?J/A+A/}}

 \author{R. Culpan,\inst{1}
 I. Pelisoli\inst{1,2}
 \and
 S. Geier\inst{1}
  } 
  \institute{Institut für Physik und Astronomie, Universität Potsdam, 
 Haus 28, Karl-Liebknecht-Str. 24/25, 14476 Potsdam-Golm, Germany
 email:rick@culpan.de
 \and
 Department of Physics, University of Warwick, Coventry, CV4 7AL, UK}

  \date{Received December 22, 2020; accepted July 27, 2021}

 
 \abstract
  {Blue horizontal-branch stars evolve from low-mass stars that have completed their main-sequence lifetimes and undergone a helium flash at the end of their red-giant phase. As such, blue horizontal-branch stars are very old objects that can be used as markers in studies of the Galactic structure and formation history. To create a clean sky catalogue of blue horizontal-branch stars, we cross-matched the Gaia data release 2 (DR2) dataset with existing reference catalogues to define selection criteria based on Gaia DR2 parameters. Following the publication of Gaia early data release 3 (EDR3), these methods were verified and subsequently applied to this latest release.}
  {Previous catalogues of blue horizontal-branch stars were developed using spectral analyses or were restricted to individual globular clusters. The purpose of this catalogue is to identify a set of blue horizontal-branch star candidates that have been selected using photometric and astrometric observations and exhibits a low contamination rate. This has been deemed important as the success of the Gaia mission has changed the way that targets are selected for large-scale spectroscopic surveys, meaning that far fewer spectra will be acquired for blue horizontal-branch stars in the future unless they are specifically targeted.}
  {We cross-matched reference blue horizontal-branch datasets with the Gaia DR2 database and defined two sets of selection criteria. Firstly, in Gaia DR2 - colour and absolute G magnitude space, and secondly, in Gaia DR2 - colour and reduced proper motion space. The main-sequence contamination in both subsets of the catalogue was reduced, at the expense of completeness, by concentrating on the Milky Way's Galactic halo, where relatively young main-sequence stars were not expected. The entire catalogue is limited to those stars with no apparent neighbours within 5 arcsec. These methods were verified and subsequently applied to the Gaia Early Data Release 3 (EDR3).}
  {We present a catalogue, based on Gaia EDR3, of 57,377 blue horizontal-branch stars. The Gaia EDR3 parallax was used in selecting 16,794 candidates and the proper motions were used to identify a further 40,583 candidates.}
  {}


  \maketitle
%

\section{Introduction}

  Blue horizontal-branch stars are old objects found on the blue side of RR Lyra at the bottom of the instability strip. They represent a late stage in the evolution of low-mass stars with initial masses in the range of ${\sim}0.8M_{\odot}$ to ${\sim}2.3\,M_{\odot}$. A star will leave the main-sequence once core hydrogen burning ceases and the inert helium core contracts. This contraction heats the hydrogen shell around the core, thereby allowing fusion to begin, and the star evolves onto the red giant branch (RGB), where a large amount of the outer envelope is shed by the strong solar wind combined with the tenuous gravitational binding forces at large radii. The inert helium core will continue to contract and becomes electron-degenerate. Finally the temperature in the helium core rises enough to allow core helium burning to begin and the helium flash occurs. This increases the temperature further and the core ceases to be degenerate. At this point, stars move to the blue horizontal-branch with masses in the range of $\sim$0.5--1.0~$M_{\odot}$ \citep{montenegro} and display helium burning in their core as well as hydrogen shell burning \citep{moehler,ruhland,paunzen}. The mass of the hydrogen shell may vary from 0.02~$M_{\odot}$ to over 0.2~$M_{\odot}$. The mass of the helium-burning core is ${\sim}$0.5~$M_{\odot}$. Blue horizontal-branch stars with a higher mass of hydrogen shell will appear cooler and, hence, redder \citep{moehler,paunzen}.
  
  The existence of horizontal-branch stars was first proven in photometric studies of the M92 and M3 globular clusters in the 1950s \citep{arp,sandage}. These observations explained the differences between the colour-magnitude diagrams from open and globular clusters first observed by Harlow Shapley in his study of the structure of the Galaxy between 1914 and 1919. Catalogues of blue horizontal-branch stars were first published in the 1980s \citep{pier,beers88} and have continued to develop as larger and higher-quality datasets have become available \citep{,beers96,beers07,montenegro}.
  
  The small scatter in their absolute magnitude and their relative abundance makes blue horizontal-branch stars a particularly useful standard candle in determining distances \citep{cleweyjarvis}. This, and the fact that they are metal-poor population II stars makes blue horizontal-branch stars ideal for studying Galactic structure in older parts of the Galaxy \citep{starkenburg,newberg,monaco,cleweyjarvis,yanny,niedersteostholt} or for studying the enclosed masses of the Milky Way within different radii \citep{sommerlarsen86,sommerlarsen89,norrishawkins,xue}. 
  
\begin{figure}
  \centering
  \includegraphics[width=\hsize]{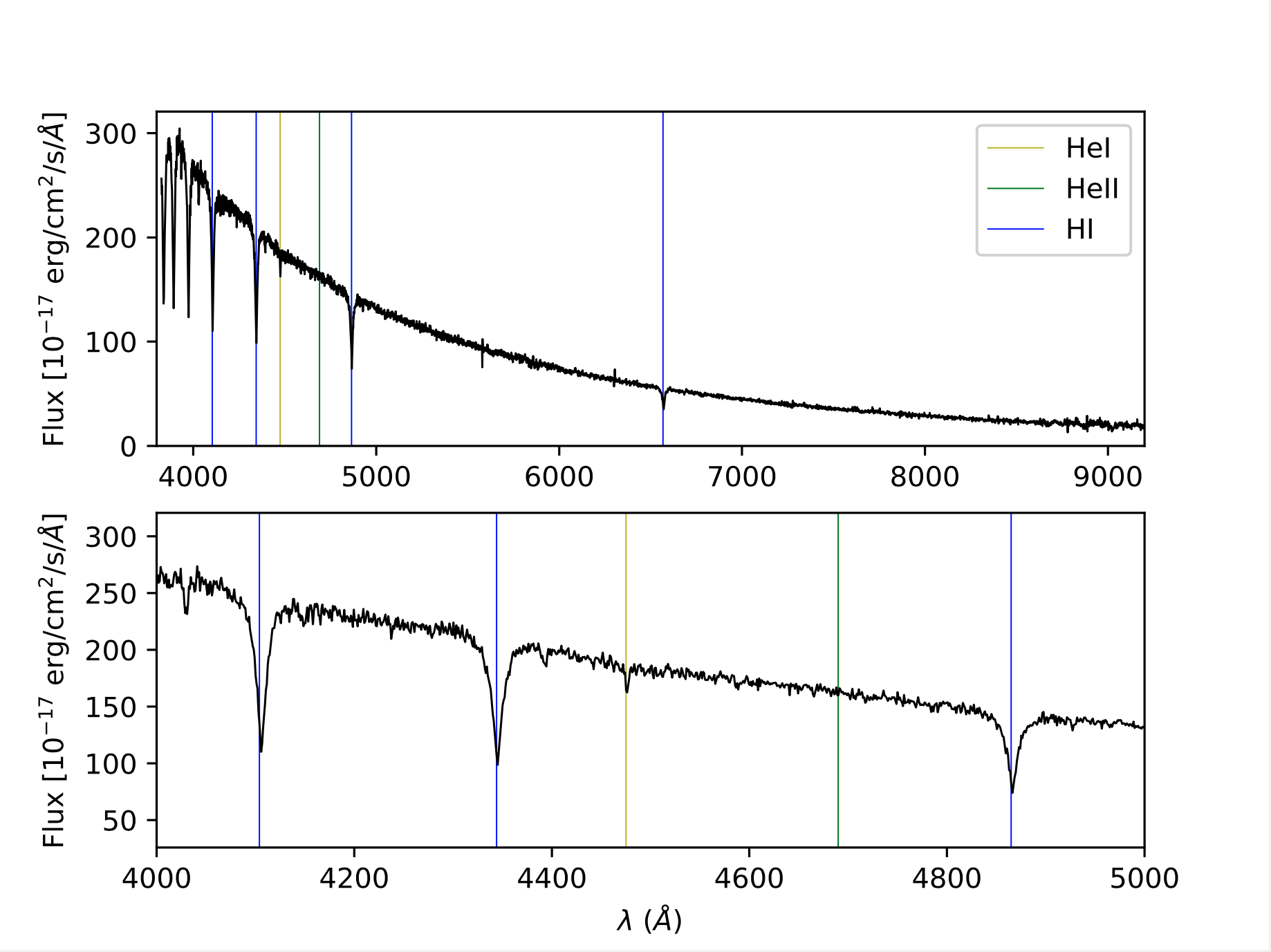}
  \includegraphics[width=\hsize]{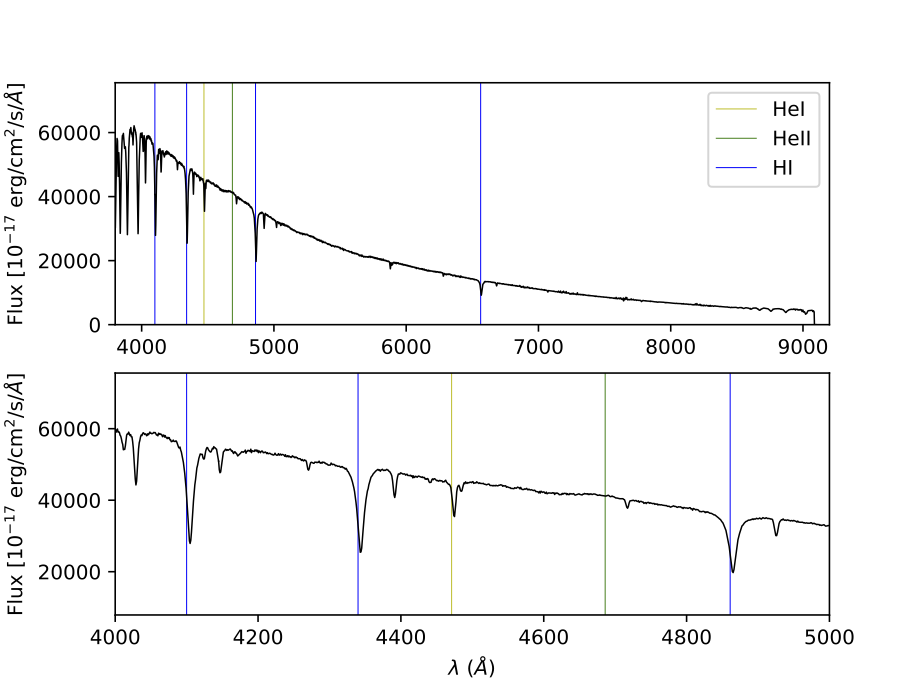}
 \caption{Typical B-type spectra displaying the difference between main-sequence and blue horizontal-branch stars. Upper panel: Example of a typical SDSS B-type spectrum from a blue- horizontal-branch star (SDSS J131916.15-011404.9 from \citet{geier_MUCHFUSS}). Lower panel: Example of a typical LAMOST B-type spectrum from a main-sequence star. Some reference lines for hydrogen and helium are indicated by vertical lines. SDSS and LAMOST spectra are comparable due to their similar resolutions and spectral coverage.}
  \label{spectra}
  \end{figure}

  Observationally, most blue horizontal-branch stars are expected to have an A-type \citep{xue} or very late B-type spectra. The A-type spectra seen in blue horizontal-branch stars can be distinguished from population I main-sequence A-type stars by their stronger Balmer jump, stronger and deeper Balmer lines, and the lack of metal spectral lines \citep{smith}. Those with an effective temperature of T$_\textrm{eff} \geq 10\,000$~K have spectra that are similar to the much younger population I B-type main-sequence stars but with weaker Helium lines, which can be distinguished spectroscopically (see Figure~\ref{spectra}). The main challenge in producing a catalogue of low contamination is differentiating between blue horizontal-branch stars and younger, population I main-sequence A-type and B-type stars as well as blue stragglers, which are older, population II main-sequence stars that have been rejuvenated through acquisition of additional mass \citep{rain} and that display A-type spectra. The additional mass may be gained by Roche Lobe overflow in a close binary system \citep{mccrea}, through stellar merger \citep{hills,davies}, or both. Blue stragglers, which have similar effective temperatures but higher gravities than blue horizontal-branch stars, are also fainter \citep{starkenburg} and are expected to be found in older parts of the Galaxy \citep{santucci}, where younger main-sequence stars of similar effective temperatures no longer exist.
  
  The goal of this work is to use the second data release (DR2) and early third data release (EDR3) of Gaia to produce a catalogue of blue horizontal-branch candidate stars with a low and quantifiable blue straggler and population I main-sequence contamination. This method identifies likely blue horizontal-branch stars without relying on spectra, thus providing important input information for upcoming large spectroscopic surveys. The blue horizontal-branch star catalogues of \citet{xue} and \citet{behr} were used as reference catalogues to calibrate the colour and absolute magnitude selection criteria in the Gaia DR2 colour magnitude space.

\section{Selection of known blue horizontal-branch stars}

The blue horizontal-branch dataset of \citet{xue} was considered the reference for determining the Gaia DR2 colour and absolute magnitude selection criteria. \citet{xue} used the seventh data release of the Sloan Digital Sky Survey \citep[SDSS DR7,][]{azerbaijan} data to obtain 2558 blue horizontal-branch stars claiming contamination of less than 10$\%$. The contaminants were considered to be main-sequence stars, particularly blue stragglers. This reference dataset was created using several steps \citep[detailed in][and references therein]{xue}. Firstly, by applying colour cuts in the $(u-g) \times (g-r)$ parameter space, then by considering the Balmer line widths at 20$\%$ below the local continuum plotted against the depth of the Balmer line compared to the local continuum, and finally by using the 'scale width versus shape' method \citep{clewley} where the Balmer lines are fit to a S\'{e}rsic profile \citep{sersic}.

The \citet{behr} dataset was used together with the \citet{xue} dataset for the purposes of comparing reference datasets with the Gaia DR2 selections outlined in subsequent sections. This was done to confirm that the selection criteria developed were valid. The \citet{behr} dataset comprises 74 blue horizontal-branch stars selected from 6 metal-poor globular clusters in the Globular Cluster Catalogue \citep{harris}.

\section{Constructing the blue horizontal-branch star catalogue from Gaia DR2}

The European Space Agency's Gaia mission was initiated with the aim of charting a three-dimensional (3D) map of the Milky Way by surveying over one billion stars in our Galaxy. This is about 1$\%$ of the stars in our Galaxy. The procedure involved creating a database of the position, parallax, brightness, colour, and proper motions of all objects visible to the satellite \citep[i.e. down to an apparent Gaia $G$ magnitude of $\sim$20.7 mag,][]{gaia16}. Each Gaia data release provides a dataset that is more complete and with reduced uncertainties than its predecessor.

Assessments of the accuracy of the astrometric and photometric measurements in the second data release \citep{gaia18a} were published by \citet{lindegren} and \citet{evans}, respectively. Assessments of the accuracy of the astrometry and photometry in the early third data release \citep{gaia20} were published by \citet{lindegren20} and \citet{riello}, respectively.
We used the TOPCAT application \citep{taylor} to cross-match the \citet{xue} blue horizontal-branch stars with the Gaia DR2 dataset to provide an initial calibration point for finding such stars in the Gaia DR2 parameter space.

The methodology used was based on the procedure used by \citet{gentilefusillo} to create a white dwarf catalogue, as well as by \citet{pelisolivos} to create a catalogue of extremely low-mass white dwarf candidates, and by \citet{geier} to create their catalogue of hot subluminous stars. Some deviation from these methodologies was required to address the challenge of main-sequence contamination as the blue horizontal-branch crosses the main-sequence leading to similar parameter values in the Gaia DR2 and EDR3 colour magnitude parameter space.

\subsection{Gaia DR2 colour and absolute magnitude criteria selection}

  \begin{figure}
  \centering
  \includegraphics[width=\hsize]{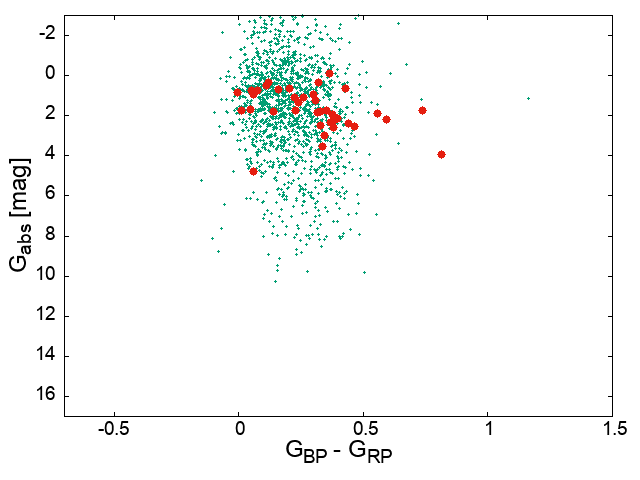}
  \includegraphics[width=\hsize]{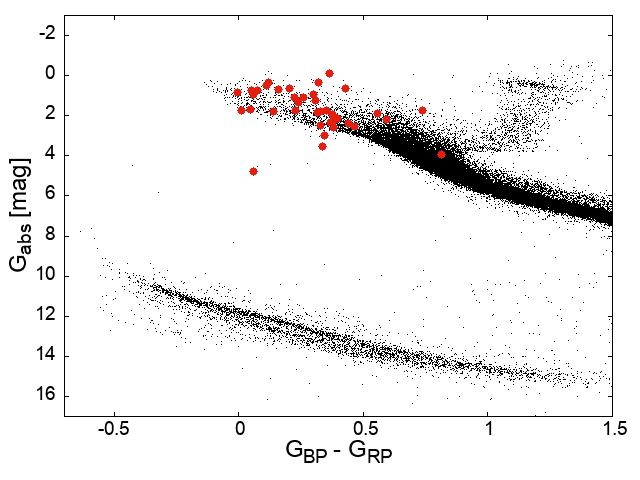}
 \caption{The result of cross-matching the \citet{xue} reference dataset with Gaia DR2. Upper panel: \citet{xue} objects plotted as turquoise dots on the Gaia DR2 colour-magnitude diagram exhibiting a large scatter in absolute magnitude due to incorrect parallax measurements. Those with parallax error <20\% are plotted as red large dots. Lower panel: \citet{lindegren} 'clean subset selection C' (black) compared with \citet{xue} objects with parallax error <20\% (red) on the same scales as the upper plot in the Gaia DR2 colour-magnitude diagram.}
  \label{hr_reference}
  \end{figure}

The \citet{xue} reference dataset was cross-matched with Gaia DR2. The absolute G magnitude was then calculated as:
\begin{center}
$G_\textrm{abs} = \verb!phot_g_mean_mag! + 5 \log_{10}(\verb!parallax!/1000) + 5$.
\end{center}
As the absolute G magnitude was calculated using the Gaia DR2 parallax measurement for distance, it was necessary to restrict the reference dataset to only those stars with a reliable parallax measurement. Using the parallax quality criterion of parallax error $<$~20$\%$ ($\verb!parallax_over_error! > 5$), as in \citet{geier} and \citet{pelisolivos}, left only 39 of the original 2560 objects from the \citet{xue} reference dataset, as shown in Figure~\ref{hr_reference}. These 39 stars were found to lie in the ranges of:
\begin{center}
$-0.1 \leq (G_{BP} - G_{RP}) \leq 0.83$ \\
$-1 \leq G_\textrm{abs} \leq 5$
\end{center}
in the Gaia DR2 colour absolute G magnitude parameter space.

These colour and absolute magnitude cut-off criteria (henceforth, referred to as the initial CMD criteria) were applied to the Gaia DR2 dataset. We also applied quality criteria that were based on the astrometry and photometry quality criteria of the 'clean subset Selection C' by \citet{lindegren}. We modified these criteria to be less stringent by allowing parallax errors up to 20$\%$ (as opposed to 10$\%$) and by relaxing the parallax lower limit to the requirement of parallax being non-negative (as opposed to being greater than 10 mas). Using the unmodified \citet{lindegren} criteria would have limited the selection to distances below $\sim100$ pc, where predominantly main-sequence Galactic disc objects are found. Furthermore, we did not apply the $\verb! astrometric_excess_noise! < 1$~mas criterion as we wanted to keep the blue horizontal-branch candidates that may exist in binary systems. As can be seen in \citet{belokurov}, in particular, Figure 7, there is an over-density of sources with a re-normalised unit weight error (RUWE) > 1 in the blue horizontal-branch region of the Gaia DR2 colour-magnitude diagram, indicating that many blue horizontal-branch candidates may exist in unresolved binary systems. \citet{lindegren} show that the astrometric excess noise to be a proxy for RUWE.

We retained the \citet{lindegren} quality criteria for the $G_{BP}$ and $G_{RP}$ photometry, which require that the errors on the $G_{BP}$ and $G_{RP}$ flux measurements are <10$\%$ and that the flux excess factor (\verb!phot_bp_rp_excess_factor!), namely:\ 
\begin{center}
$E = (I_{BP} + I_{RP}) / I_G$
\end{center}
lies in the range defined by the following cut-offs:
\begin{center}
$\verb!phot_bp_rp_excess_factor! > 1.0 + 0.015(G_{BP} - G_{RP})^2,$ \\
$\verb!phot_bp_rp_excess_factor! < 1.3 + 0.06(G_{BP} - G_{RP})^2.$ \\
\end{center}

Applying the initial CMD criteria and the modified \citet{lindegren} criteria for astrometric and photometric quality to Gaia DR2 resulted in a set of over 9 million objects. These stars are expected, based on their position on the colour-magnitude diagram (CMD), to be blue horizontal-branch stars, blue-stragglers, and population I main-sequence A-type and B-type stars. The measured parallaxes of these objects lead us to believe that they exist in either the Galactic disc or the Galactic halo. 

It is known that the relatively young population I main-sequence A-type and B-type stars are not expected in the much older Galactic halo. We expected that focusing on the Galactic halo will reduce the population I main-sequence contamination of the dataset. To do this we applied the stringent halo selection criteria ($V_T > 200$~km s$^{-1}$, $|b|>60^\circ$) to the 9 million objects found so far. Using a tangential velocity cut-off to differentiate between disk and halo objects is an accepted method \citep{gaia18,koppelman} that uses only the two tangential components of velocity. It should be noted that any halo objects with low tangential velocities but with high radial velocities along the line of sight will not be selected. This is more likely to be the case in the direction of the motion of the heliocentric system within the Galactic reference frame. We calculated the tangential velocity as:
\begin{center}
 $V_T = (4.74/\verb!parallax!) \times (\verb!pmra!^2 + \verb!pmdec!^2)^{1/2}$.
\end{center}

\begin{figure*}
  \centering
  \includegraphics[width=0.9\hsize]{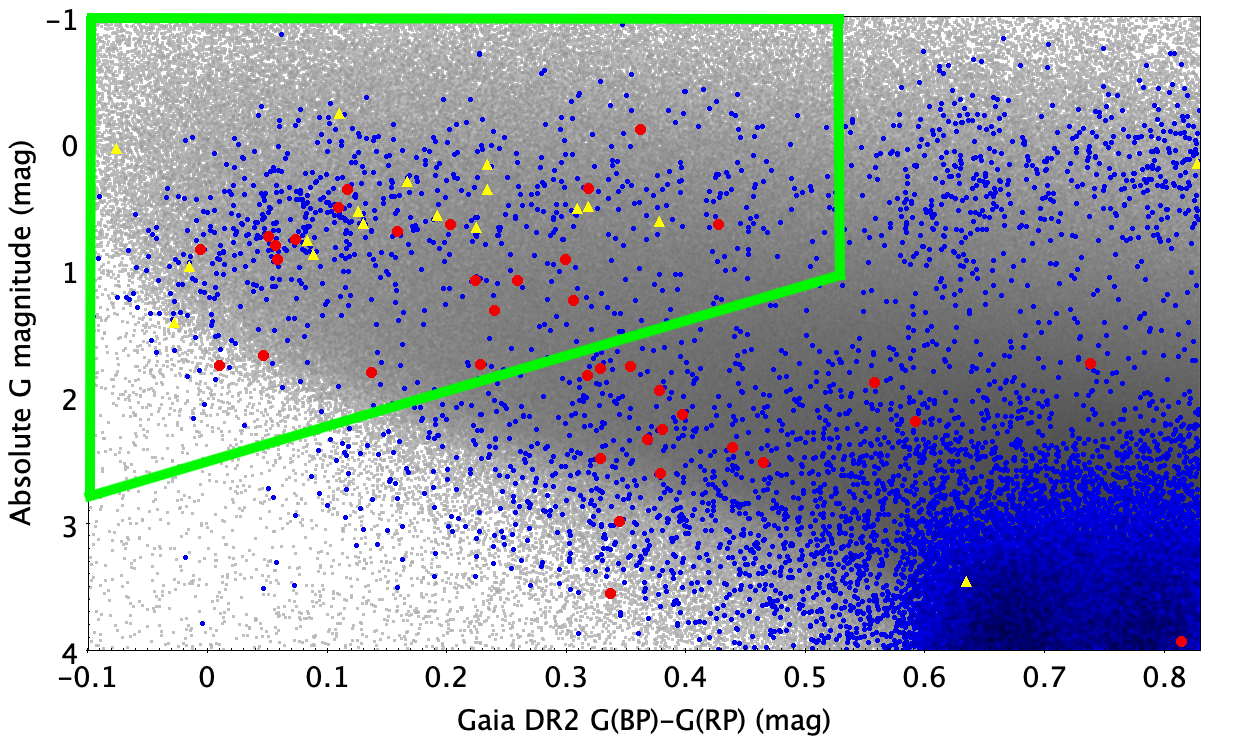}
  \includegraphics[width=0.9\hsize]{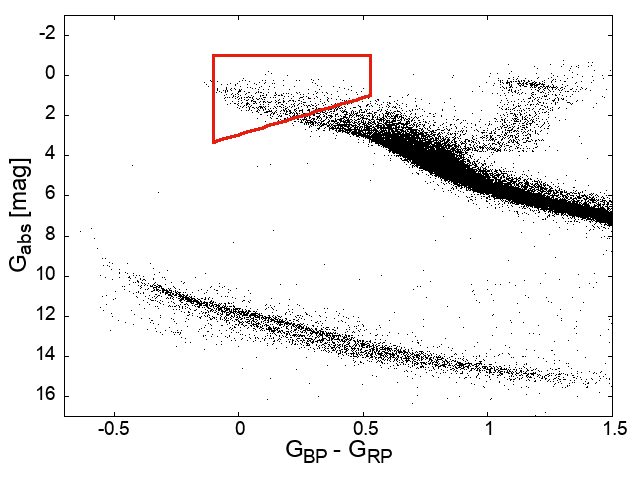}
 \caption{Gaia DR2 colour absolute magnitude diagram showing the blue horizontal-branch star region. Top panel: Gaia DR2 objects that satisfy the modified \citep{lindegren} criteria (gray points) with those with the stringent halo criteria applied (small blue dots). \citet{xue} objects (red circles) and \citet{behr} objects (yellow triangles) are superimposed. The green trapezium denotes the Gaia DR2 colour-magnitude coordinate space (final CMD criteria) considered as the blue horizontal-branch region. The main-sequence turn-off can be seen at around $(G_{BP} - G_{RP})$ = 0.6 and $G_{abs}$ = 3. Blue stragglers are found extending the main-sequence beyond that. Our selection was designed to minimise the blue-straggler contamination and focus on the blue horizontal-branch stars. Bottom panel: For comparison, the \citet{lindegren} 'clean subset selection C' comprised of disc stars within 100pc which are primarily main-sequence stars (black) with red trapezium denoting the Gaia DR2 colour-magnitude coordinate space considered as the blue horizontal-branch region.}
  \label{hr_parallax}
  \end{figure*}

This resulted in a set of 43,040 objects that, when plotted on the Gaia DR2 CMD that showed a clear clustering in the region where blue horizontal-branch stars would be expected. This was confirmed by overlaying the reference datasets from \citet{xue} and \citet{behr} (see Figure~\ref{hr_parallax}).

We were then able to define modified CMD selection criteria that would allow us to focus more closely on the blue horizontal-branch region in Gaia DR2 colour-magnitude space. We defined these modified criteria as:

\begin{center}
$ -0.1 < (G_{BP} - G_{RP}) < 0.53$,
$-1 < G_{abs} < (2.5 - 2.77(G_{BP} - G_{RP})),$
\end{center}

henceforth the final Gaia CMD criteria.
Applying the final CMD selection criteria to the 9 million initial CMD objects reduced the selection to 305,622 objects comprising of blue horizontal-branch stars, blue-stragglers, and population I main-sequence A-type and B-type stars. It is known that regions of high main-sequence stellar population density and hence, high potential population I main-sequence contamination exist in the Galactic plane and the Magellanic Clouds. These high apparent stellar density regions are also prone to having astrometric solutions that are not accurate, but still precise, and therefore not easily removed by a quality assessment alone (e.g. \citet{gentilefusillo}). Objects in these regions were removed by excluding regions in the Galactic plane and the Magellanic Clouds with a local apparent stellar population density $>$50,000 per deg$^2$:
\begin{center}
$-25^{\circ} < b < 25^{\circ}$,
\end{center}
the Large Magellanic Cloud:
\begin{center}
$274.5^{\circ} < l < 286.5^{\circ}$ and $-37.9^{\circ} < b < -27.9^{\circ}$,
\end{center}
and the Small Magellanic Cloud:
\begin{center}
$299.8^{\circ} < l < 305.8^{\circ}$ and $-46.3^{\circ} < b < -42.3^{\circ}$.
\end{center}

\begin{figure*}
  \centering
  \includegraphics[width=\hsize]{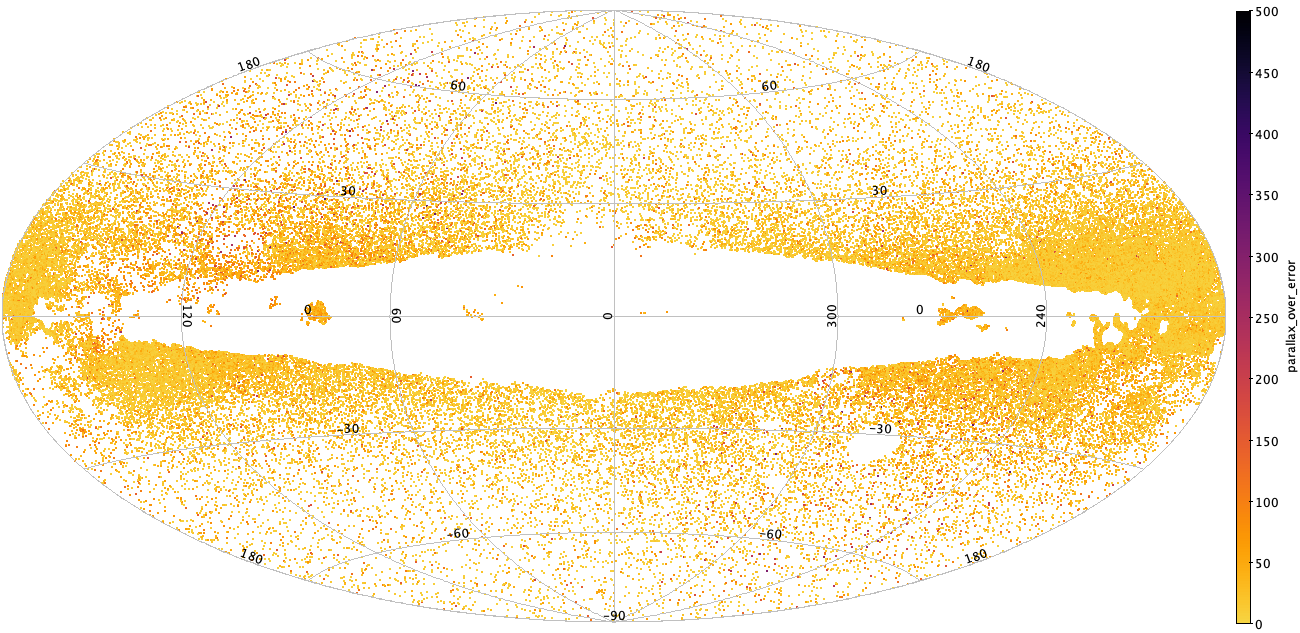}
 \caption{Sky distribution of parallax selection candidates. Regions of high stellar number density (>50,000 stars per square degree) have been excluded.}
  \label{bhb_parallax_dist}
  \end{figure*}

This further reduced the number of objects in our selection to 104,763.
Fainter objects in crowded regions are known to be susceptible to photometric errors in the $G_{BP}$ and $G_{RP}$ bands \citep{lindegren}. Thus, additional filtering was added whereby stars with apparent neighbours within 5 arcsec were rejected \citep{pelisolivos, gentilefusillo, boubert}. This close apparent neighbour filtering criterion had, together with the Galactic plane and Magellanic Cloud filtering, the additional benefit of removing the majority of objects where reddening from extinction in dust clouds is not negligible. We did not apply extinction corrections to the Gaia DR2 colours for these objects due to the low resolution of existing dust maps \citep{pelisolivos}. Removing any remaining objects with close apparent neighbours further reduced the number of selected objects to 91,719 which are expected to comprise of blue horizontal-branch stars, blue stragglers, and a reduced proportion of population I main-sequence A-type and B-type stars (see Figure~\ref{bhb_parallax_dist}).

\begin{table*}[h!]
\centering
\begin{tabular}{c} 
 \hline
 \hline
 Initial {\it Gaia} CMD criteria: \\
 \hline
 $-0.1 \leq (G_{BP} - G_{RP}) \leq 0.83$ \\
 $-1 \leq G_{abs} \leq 5$ \\ [0.5ex]
 \hline
 \hline
 Final {\it Gaia} CMD criteria: \\
 \hline
 $ -0.1 < (G_{BP} - G_{RP}) < 0.53$ \\
 $-1 < G_{abs} < (2.5 - 2.77(G_{BP} - G_{RP}))$ \\ [0.5ex]
 \hline
 \hline
 Modified \citet{lindegren} criteria for astrometric quality: \\
 \hline
 $\verb!parallax! > 0$ \\ 
 $\verb!parallax_over_error! > 5$ \\ [0.5ex]
 \hline
 \hline
 Modified \citet{lindegren} criteria for photometric quality: \\
 \hline
 $\verb!phot_bp_mean_flux_over_error! > 10$ \\
 $\verb!phot_rp_mean_flux_over_error! > 10$ \\
 $\verb!phot_bp_rp_excess_factor! > 1.0 + 0.015(G_{BP} - G_{RP})^2$ \\
 $\verb!phot_bp_rp_excess_factor! < 1.3 + 0.06(G_{BP} - G_{RP})^2$ \\ [0.5ex]
 \hline
 \hline
 Stringent halo selection criteria: \\
 \hline
 $V_T > 200$~km s$^{-1}$ \\
 $|b|>60^\circ$ \\ [0.5ex]
 \hline
 \hline
 Crowded region criteria of \\
 $>$50,000 stars within 1 deg$^2$ for: \\
 \hline
 Galactic plane: \\
 ($-25^{\circ} < b < 25^{\circ}$) \\
 \hline
 Large Magellanic Cloud: \\
 $274.5^{\circ} < l < 286.5^{\circ}$ and $-37.9^{\circ} < b < -27.9^{\circ}$ \\
 \hline
 Small Magellanic Cloud: \\
 $299.8^{\circ} < l < 305.8^{\circ}$ and $-46.3^{\circ} < b < -42.3^{\circ}$ \\ [0.5ex]
 \hline
 \hline
 Close apparent neighbour criterion: \\
 \hline
 no apparent neighbour within 5 arcsec \\ [0.5ex]
 \hline
 \hline
\end{tabular}
\caption{Table of selection criteria applied to Gaia DR2 to define blue horizontal-branch star candidate selection from colour and absolute G magnitude (parallax selection)}
\label{table:1}
\end{table*}

\subsubsection{Galactic halo selection}

The stringent halo selection criteria were relaxed in small iterative increments in order to find the least restrictive cut-offs before a disproportionate increase in the number of candidate objects was observed indicating population I main-sequence and non-halo object contamination. A cut-off for Galactic latitude was not applied and a value of tangential velocity of $V_T > 145$~km s$^{-1}$ was found for the parallax candidate objects. This was considered the Galactic halo selection criterion for the parallax error $<$~20$\%$ dataset (see Table 2) and was applied to the parallax candidates to produce a cleaner set of halo, blue horizontal-branch stars and, thus, a cleaner selection of parallax candidates.

\subsubsection{Verification of the Galactic halo selection using LAMOST spectra}

The 91,719 objects found prior to applying any halo selection criteria were cross-matched with the SDSS data release 16 \citep[DR16,][]{ahumada} and LAMOST fifth data release (DR5) \citep{yao} surveys to find spectra for the blue horizontal-branch candidate objects. The relatively close proximity of the parallax candidates leads to their being brighter than more distant candidates. The low apparent magnitudes (lower than the SDSS DR16 photometric saturation limit) of the objects being cross-matched with the SDSS DR16 spectra meant that less than 100 matches were found which we considered to be not statistically significant and thus, may not be representative of the dataset. LAMOST DR5 spectra were, however, found for $\sim$24,000 stars. This represented roughly one third of the objects found had a similar magnitude distribution and were thus considered to be a statistically significant and representative population.

The LAMOST stellar parameter pipeline \citep{xiang} found that $\sim$21,500 stars had A-type spectra, as would be expected for most blue horizontal-branch stars, for blue stragglers, and for population I main-sequence A-type stars \citep{xue}. Furthermore, about 880 were found to have B-type spectra, as would be expected for hotter blue horizontal-branch stars (see the upper panel of Figure~\ref{spectra} for a typical example) and population I main-sequence B-type stars (see the lower panel of Figure~\ref{spectra} for a typical example). The remaining $\sim$1600 spectra were considered to be too noisy for a spectral classification. Visual inspection of these spectra confirmed the LAMOST classification for the A-type and B-type spectra and less than 1\% of the LAMOST objects were manually re-categorised.

The spectra of objects categorised by the LAMOST stellar parameter pipeline as being B-type stars were found to have distinct helium lines as would be expected for population I main-sequence B stars. Blue horizontal-branch B-type stars display much subtler helium lines. Thus, we used the presence of large numbers of LAMOST B-type spectra as a proxy for the presence of population I main-sequence stars and, hence, as a proxy for non-halo objects.

The application of the halo selection criterion of $V_T > 145$~km s$^{-1}$ reduced the number candidate objects by around 90\% while reducing the number of LAMOST cross-matched objects with B-type spectra from $\sim$880 to 6, a reduction of over 99\%. This represents an effective filter with which we can discern between disc and halo objects, thereby facilitating the removal of a large potential source of population I main-sequence contamination.

\subsection{Gaia reduced proper motion criteria selection}

Selecting only the Gaia DR2 stars with parallax errors of $<$~20$\%$ results in stars that are more than a few hundred pc away being rejected. New selection criteria that do not use the Gaia parallax measurement needed to be found in order to find candidate stars in this large population of more distant objects. The parallax was used in calculating the absolute magnitude and the tangential velocity in the parallax\_error < 20$\%$ dataset.

Reduced proper motion (H) is calculated from total proper motion (\(\mu\)) and can be used as a proxy for absolute magnitude \citep{gentilefusillo2015,geier}:
\begin{center}
$\mu = (\verb!pmra!^2 + \verb!pmdec!^2)^{1/2}$, \\ [0.5ex]
$H = {\verb!phot_g_mean_mag!} + 5 + (5(log_{10}(\mu / 1000)),$ \\ [0.5ex]
$\sigma_{\mu} = (\verb!pmra_error!^2 + \verb!pmdec_error!^2)^{1/2}$, \\ [0.5ex]
${\verb!total_proper_motion_over_error!} = \mu/\sigma_{\mu} $.
\end{center}

  \begin{figure}
  \centering
  \includegraphics[width=\hsize]{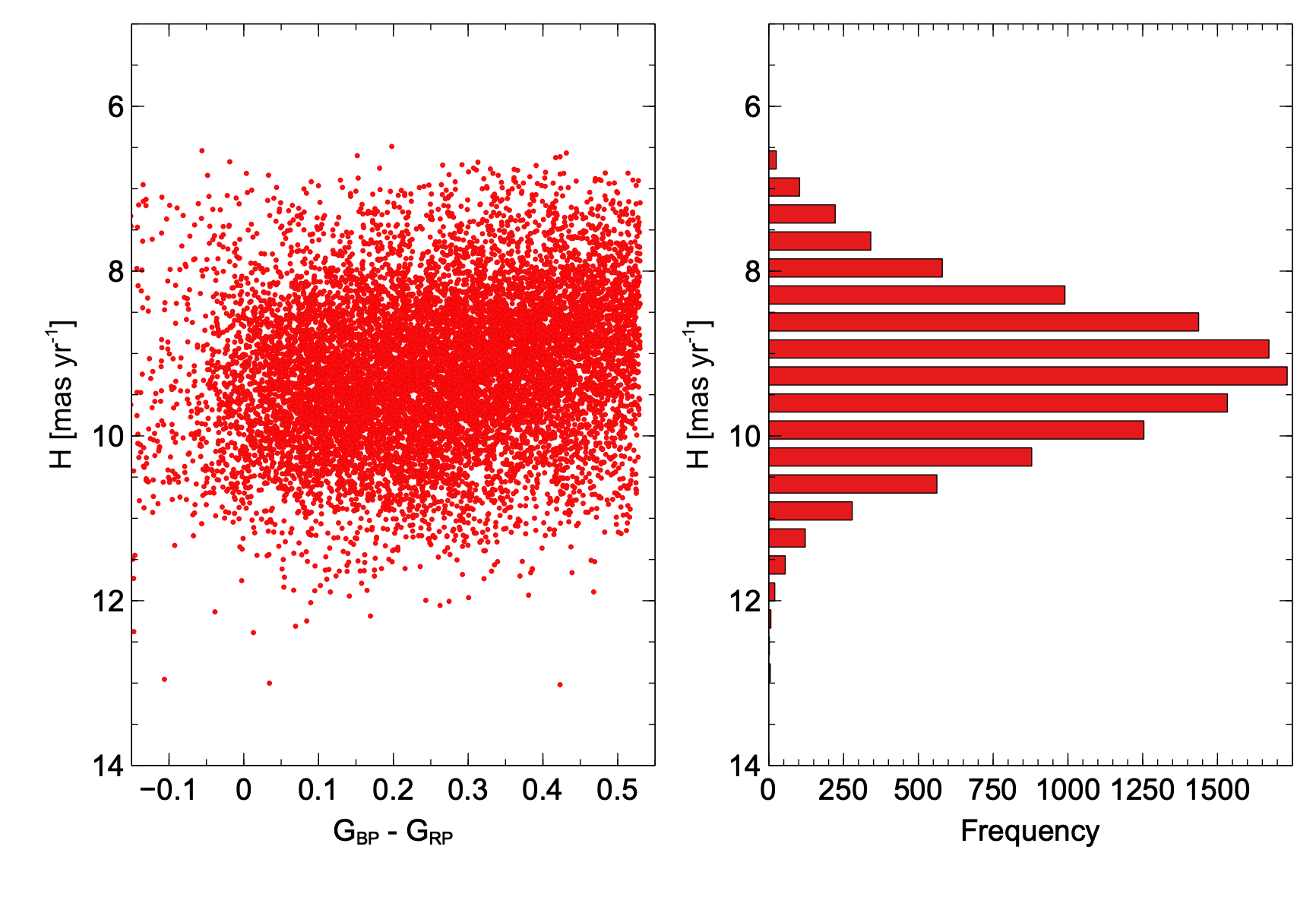}
 \caption{The Gaia DR2 reduced proper motion versus colour diagram and reduced proper motion frequency plot used to define the selection criteria for the objects with a parallax error $\geq$~20$\%$. Left panel: Reduced proper motion (H) versus colour (\(G_{BP}-G_{RP})\) diagram for the blue horizontal-branch candidates with parallax error $<$~20$\%$. Right panel: Distribution of reduced proper motion (H) seen in the blue horizontal-branch candidates with a parallax error < \(20\%\).}
  \label{redpm_for_parallax}
  \end{figure}

Figure~\ref{redpm_for_parallax} shows the distribution of reduced proper motions seen in the objects from the parallax selection. This was used to define the selection criteria for the blue horizontal-branch candidates from the parallax error $\geq$~20$\%$ objects (henceforth, called the proper motion candidates).

The {\verb!total_proper_motion_over_error!} was also calculated in order to provide a quality criterion to prevent stars with large proper motion uncertainties being used \citep{gentilefusillo2015}. This is analogous to using {\verb!parallax_over_error!} as a quality criterion in the parallax error $<$~20$\%$ dataset. A quality cut-off of total proper motion error < 20\% was used in line with the parallax error $<$~20$\%$ cut-off used previously.

The plot of Gaia DR2 colour versus reduced proper motion for the parallax candidates dataset (Figure~\ref{redpm_for_parallax}) gave a distribution of reduced proper motions with $\overline{H} = 9.36, \sigma_{H} = 0.81$. A range of $\pm3\sigma$ about the mean reduced proper motion gave a 99$\%$ certainty of capturing all blue horizontal-branch candidates using reduced proper motion as a proxy for absolute magnitude.

The same photometric quality criteria as used in selecting the parallax candidate dataset were also applied for the proper motion candidate dataset. The astrometric quality criteria used for the parallax candidate dataset were removed but the criterion $\verb!parallax_over_error! \leq 5$ was applied to ensure that no stars were to be found in both datasets.

Application of the total proper motion over error cut-off, the reduced proper motion and colour criteria, the photometric quality criteria, crowded region criterion, the Galactic latitude criterion, and the parallax error cut-off to the Gaia DR2 dataset resulted in the proper motion selection dataset.

\begin{figure}
  \centering
  \includegraphics[width=\hsize]{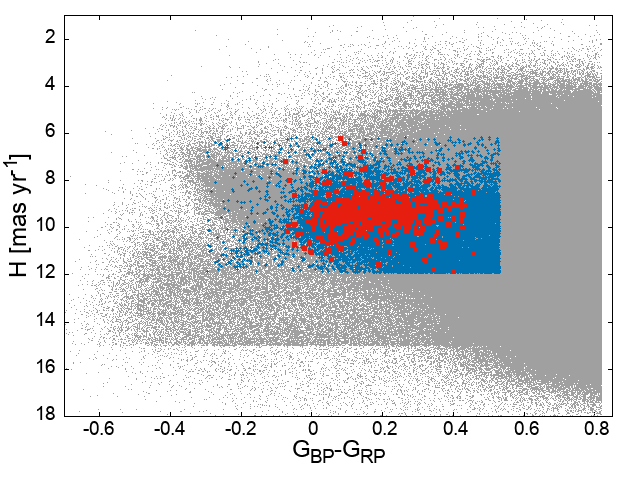}
 \caption{Gaia DR2 colour and reduced proper motion plot:\ Objects conforming to the crowded region criterion as well as the total proper motion and photometric quality criteria (grey dots), objects also conforming to the blue horizontal-branch and halo criteria (blue crosses), and \citet{xue} objects (red squares).}
  \label{gaia_colour_redpm}
  \end{figure}

Figure~\ref{gaia_colour_redpm} shows the Gaia DR2 objects that conform to the total proper motion over error cut-off, the photometric quality and the crowded region criteria, the subset that conforms to the blue horizontal-branch and halo criteria with the reference objects \citep{xue} overlain. 

\subsubsection{Galactic halo selection}

As the parallax is used in calculating the tangential velocity this could no longer be reliably used as the halo selection criterion. We used the Galactic latitude for this purpose. As done previously, we applied the stringent halo selection criterion (absolute Galactic latitude $|b|>60^\circ$) and we relaxed this criterion in small iterative increments in order to find the least stringent cut-offs before a disproportionate increase in the number of candidate objects was observed indicating population I main-sequence and non-halo object contamination. The halo selection criterion found was $|b|>50^\circ$.

\subsubsection{Verification of the Galactic halo selection using LAMOST spectra}

As previously done for the parallax selection, the halo selection criterion was verified by comparing the number of LAMOST B-spectra removed to the total number of objects removed. Application of the halo selection criterion of $|b|>50^\circ$ to the $\sim$24,000 LAMOST cross-matched stars, presented in Section 3.1.2, reduced the number candidate objects by around 92\%, while reducing the number of LAMOST cross-matched objects with B-type spectra from $\sim$880 to 2, a reduction of over 99.8\%. This represents an effective filter for discerning between disc and halo objects, thereby facilitating the removal of a large potential source of population I main-sequence contamination.


\section{Application of the Gaia GR2 selection criteria to Gaia EDR3}

Upon the release of Gaia EDR3, the processing steps outlined in the sections above were performed using Gaia EDR3 and the results were compared to those achieved with Gaia DR2. Firstly, we did this to verify that the methodology and criteria developed for the Gaia DR2 dataset was still applicable in the latest release. Secondly, this was done to create a catalogue of blue horizontal-branch stars using Gaia EDR3.

\begin{figure}
  \centering
  \includegraphics[width=\hsize]{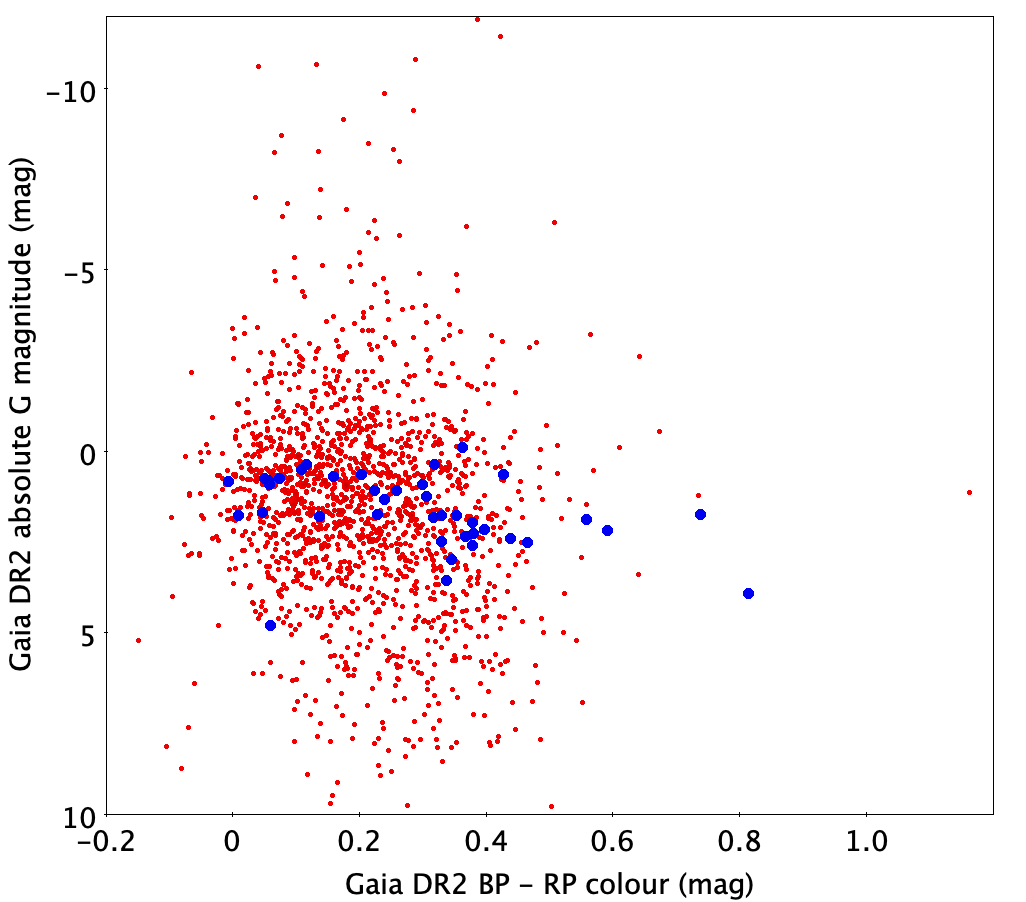}
  \includegraphics[width=\hsize]{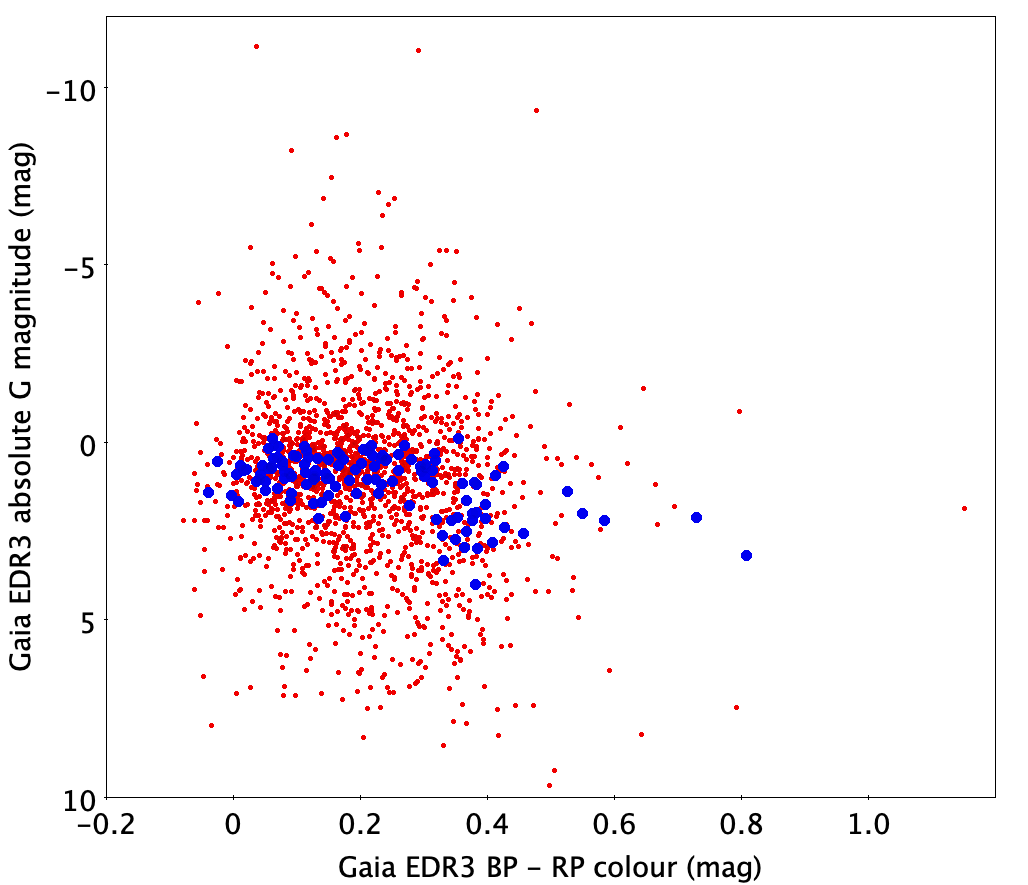}
 \caption{\citet{xue} blue horizontal-branch reference objects (red dots) displayed in DR2 (upper panel) and EDR3 (lower panel). Those objects with a parallax error of less than 20$\%$ are shown as blue squares.}
  \label{xue_dr2_edr3}
  \end{figure}

The 2558 blue horizontal-branch reference objects from \citet{xue} were found in Gaia EDR3 absolute magnitude-colour space. The improved accuracy of the parallax measurements compared to Gaia DR2 meant that 121 objects were found to have a parallax error of less than 20$\%$ in Gaia EDR3 (compared to 39 in Gaia DR2). See Figure~\ref{xue_dr2_edr3}. The position of these reference objects in the Gaia EDR3 colour and absolute magnitude space verified that using the same colour and absolute magnitude criteria that were found using Gaia DR2 was valid.

\begin{figure}
  \centering
  \includegraphics[width=\hsize]{GaiaDR2CMD_BHB_Halo_Xue_Behr.png}
  \includegraphics[width=\hsize]{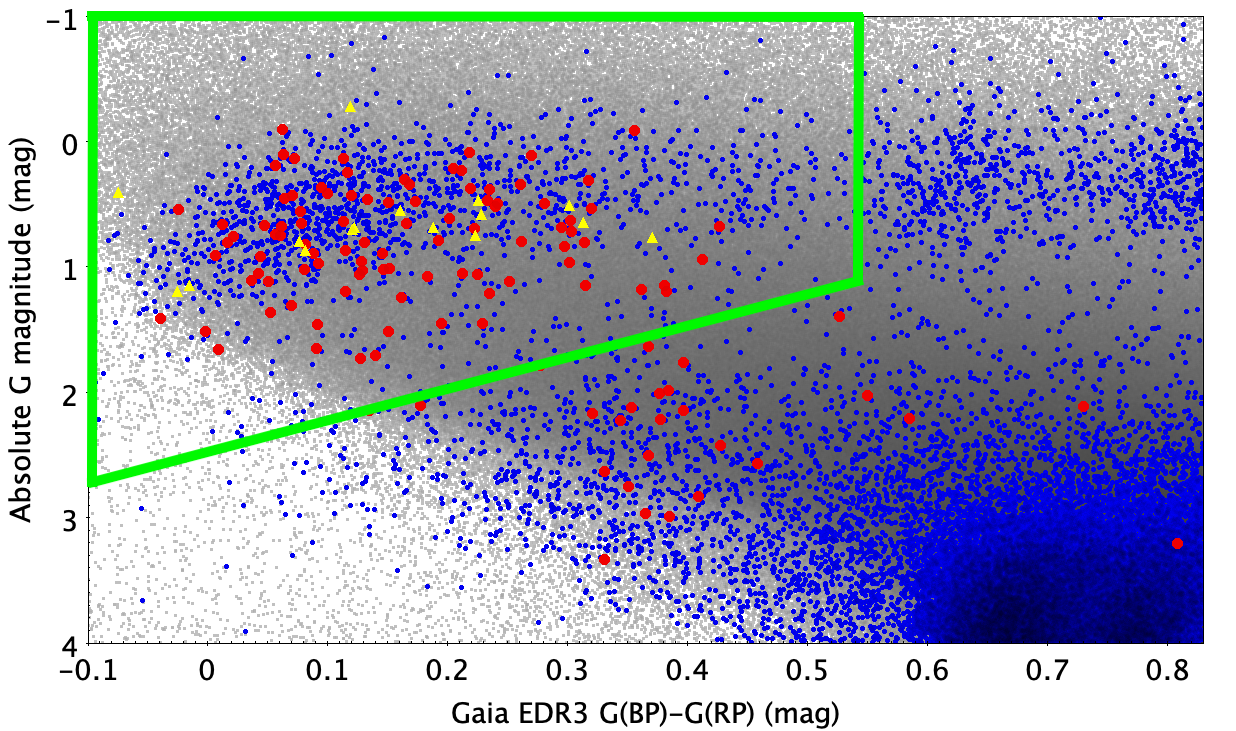}
 \caption{Top panel: Objects that satisfy the modified \citep{lindegren} criteria (gray points) with those with the stringent halo criteria and the crowded region criterion applied (small light blue dots). \citet{xue} objects (red circles) and \citet{behr} objects (yellow triangles) are superimposed.The green trapezium denotes the Gaia DR2 colour-magnitude coordinate space (final CMD criteria) considered as the blue horizontal-branch region.) displayed in DR2 (upper panel) and EDR3 (lower panel).}
  \label{CMD_dr2_edr3}
  \end{figure}

Thus, the same selection criteria were applied to Gaia EDR3 and the same stringent halo selection criteria were applied (see Figure~\ref{CMD_dr2_edr3}). We inspected the results and found that the final DR2 CMD criteria as defined in Section 3.1
\begin{center}
$-0.1 < (G_{BP} - G_{RP}) < 0.53$; \\
$-1 < G_\textrm{abs} < (2.5 - 2.77(G_{BP} - G_{RP}))$
\end{center}
were also applicable to EDR3. 

  \begin{figure}
  \centering
  \includegraphics[width=\hsize]{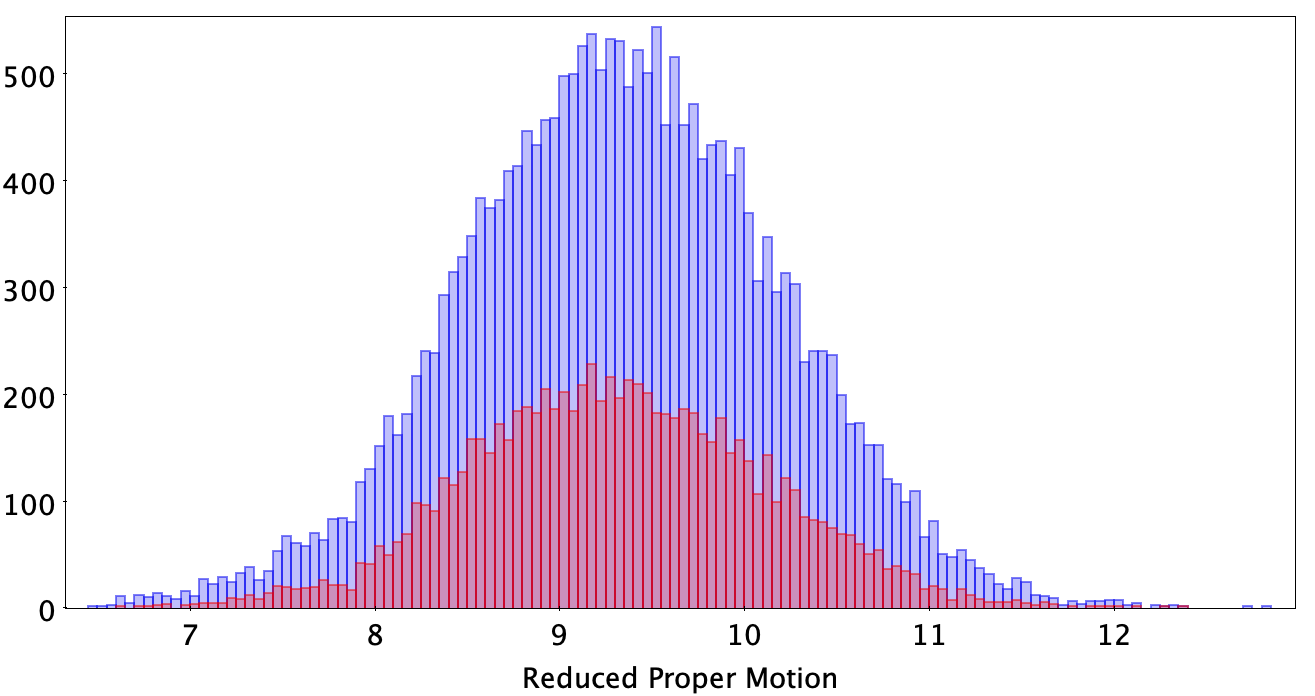}
 \caption{Distribution of reduced proper motions for the parallax selection in Gaia DR2 (red) and Gaia EDR3 (blue).}
  \label{RedPM_Dist_dr2_edr3}
  \end{figure}

The observed reduced proper motion distribution from Gaia EDR3 parallax selection was then used in defining the proper motion candidate selection criteria. We compared the reduced proper motion distribution of these objects from the Gaia DR2 and Gaia EDR3 datasets and found a close agreement despite the larger number of objects found in Gaia EDR3 (see Figure~\ref{RedPM_Dist_dr2_edr3}).

\subsection{Comparison of Gaia DR2 and EDR3 results}

\begin{table*}[h!]
\centering
\begin{tabular}{ l c c }
\hline\hline
Selection Criteria & DR2 & EDR3 \\ [0.5ex]
\hline
1. Parallax error < 20$\%$, initial Gaia CMD Criteria & 9,019,816 & 10,921,642 \\
2. Position 1 + final Gaia CMD Criteria & 305,622 & 336,911 \\
3. Position 2 + crowded region criterion & 104,763 & 121,669 \\
4. Position 3 + no apparent neighbour within 5 arcsec criterion & 91,719 & 101,913 \\
5. Position 4 + $V_T > 145$~km s$^{-1}$ & 8,556 & 16,794 \\
6. Reduced proper motion selection & $ 6.94 < H < 11.78$ & $6.94 < H < 11.81$ \\
7. Position 6 + parallax error >= 20$\%$ + $|b|>50^\circ$ & 31,957 & 42,091 \\
8. Position 7 + no apparent neighbour within 5 arcsec criterion & 30,919 & 40,583 \\
\hline\hline
\end{tabular}
\caption{Table showing a comparison of the Gaia DR2 and EDR3 selection criteria and results}
\label{table:2}
\end{table*}

We found that 7,307 of the 8,556 objects in the Gaia DR2 parallax selection were also present in the Gaia EDR3 parallax selection. The 1,249 parallax candidate objects that are in the DR2 dataset but no longer in the EDR3 dataset are found on the Gaia CMD with increasing concentration towards the $-1 < G_{abs} < (2.5 - 2.77(G_{BP} - G_{RP}))$ cut-off criterion. This criterion removes millions of objects, some of which will have measured values that deviate from their true value by multiple standard deviations causing an increase in contamination towards this highly populated region in the Gaia colour-magnitude space.

\begin{figure}
  \centering
  \includegraphics[width=\hsize]{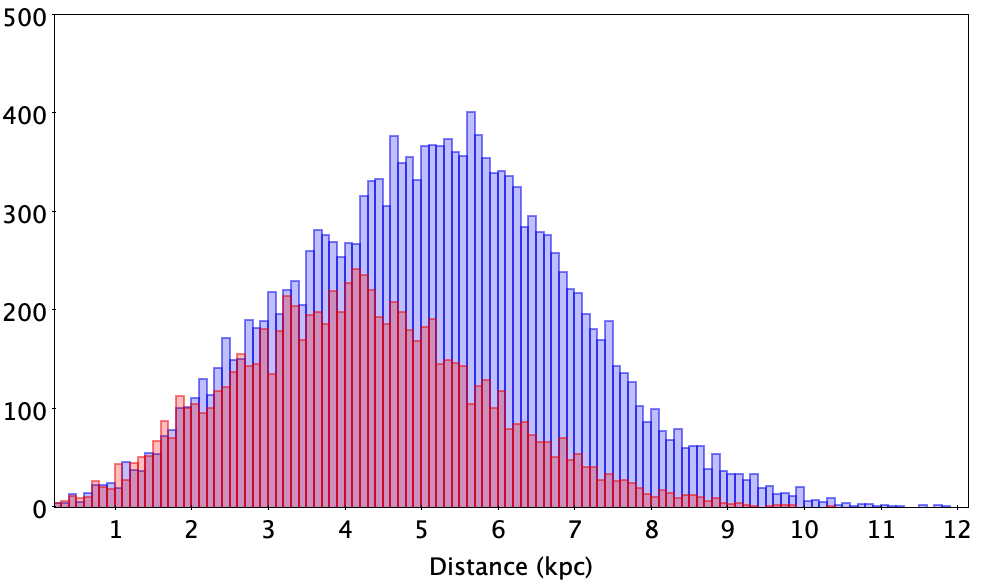}
 \caption{Distance distribution of parallax selection objects in Gaia DR2 (red) and Gaia EDR3 (blue).}
  \label{Dist_Dist_dr2_edr3}
  \end{figure}

By comparing the distance (calculated as the reciprocal of parallax) distributions of the parallax selections from Gaia DR2 and Gaia EDR3, we can demonstrate that we have been able to identify more distant candidate objects associated with higher apparent magnitudes (see Figure~\ref{Dist_Dist_dr2_edr3}).

\begin{figure}
  \centering
  \includegraphics[width=\hsize]{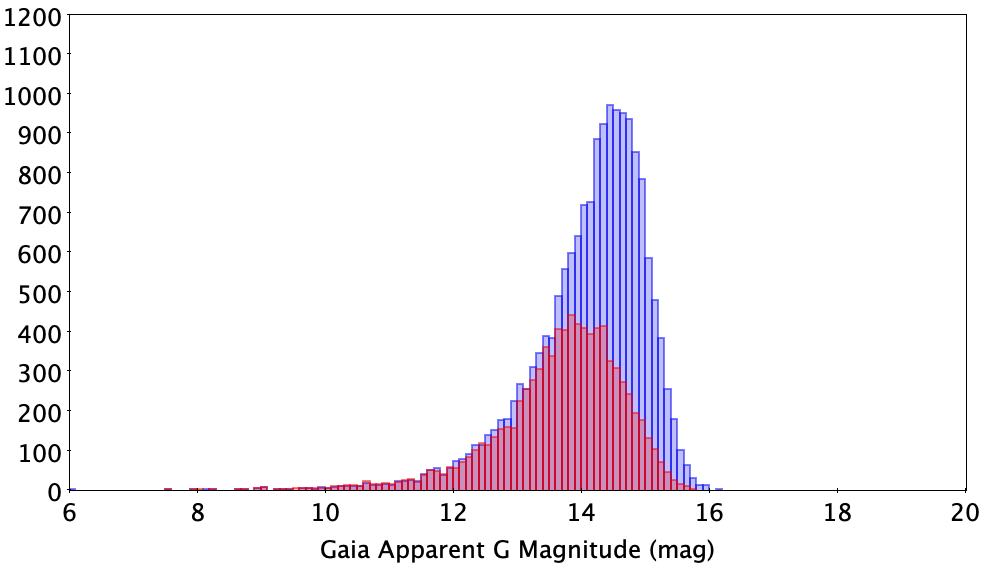}
  \includegraphics[width=\hsize]{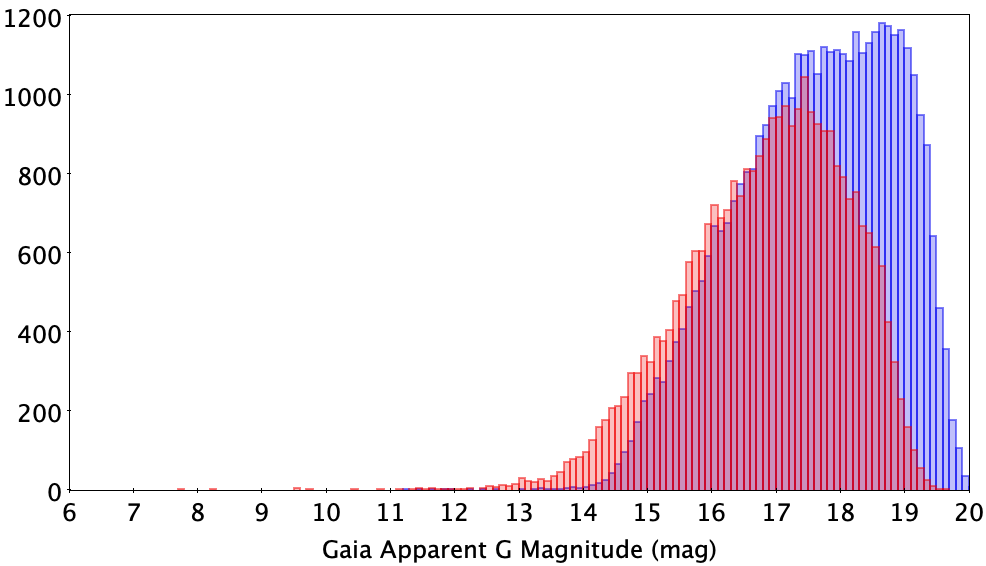}
 \caption{Distribution of apparent magnitudes for the parallax selection (upper panel) for Gaia DR2 (red) and Gaia EDR3 (blue). Distribution of apparent magnitudes for proper motion selection (lower panel) in Gaia DR2 (red) and Gaia EDR3 (blue).}
  \label{AppMag_Dist_dr2_edr3}
  \end{figure}

Making such a comparison for objects with high uncertainties on their parallaxes may be misleading. As blue horizontal-branch stars are considered to have nearly constant absolute magnitudes we can consider apparent magnitude as a proxy for distance. The comparison of Gaia DR2 and Gaia EDR3 catalogues can be meaningfully made in both the parallax selection and the proper motions selection in Figure~\ref{AppMag_Dist_dr2_edr3}.

We compared the proper motions candidate objects found in Gaia DR2 and the parallax candidate objects found in Gaia EDR3. We observed that 1,737 objects that were found in the Gaia DR2 proper motions catalogue were now to be found in the Gaia EDR3 parallax dataset. Our examination of the apparent magnitudes of these 1,737 objects showed them to be at the fainter (average apparent G magnitude 14.38 mag) and, hence, at the more distant end of the dataset. Thus, the increased accuracy of the parallax measurements made in Gaia EDR3 enables the identification of halo blue horizontal-branch objects over the full sky out to a greater distance than was previously possible using these methods in Gaia DR2.

\section{Estimation of catalogue contamination and completeness}

We made an estimation of the population I main-sequence and blue-straggler contamination levels using the SDSS SEGUE Stellar Parameter Pipeline (SSPP) DR10 \citep{lee,smolinski}, based on its effective temperature ($T_{eff}$) and surface gravity (log g) parameters . The method used and the cut-offs applied were taken from Santucci et al. (2015 and references therein), as noted in Section 3.1.1.

\begin{figure}
  \centering
  \includegraphics[width=\hsize]{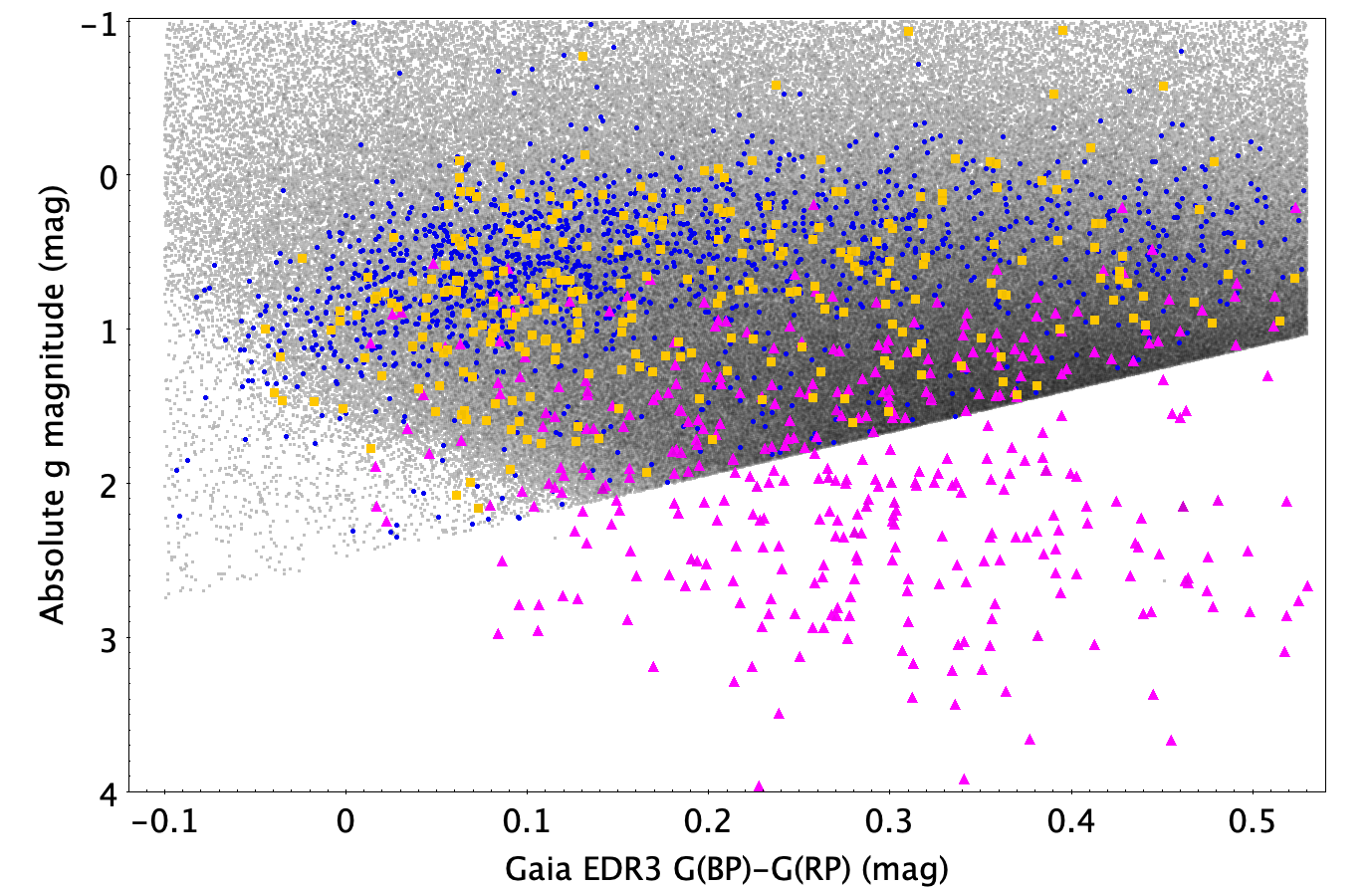}
  \includegraphics[width=\hsize]{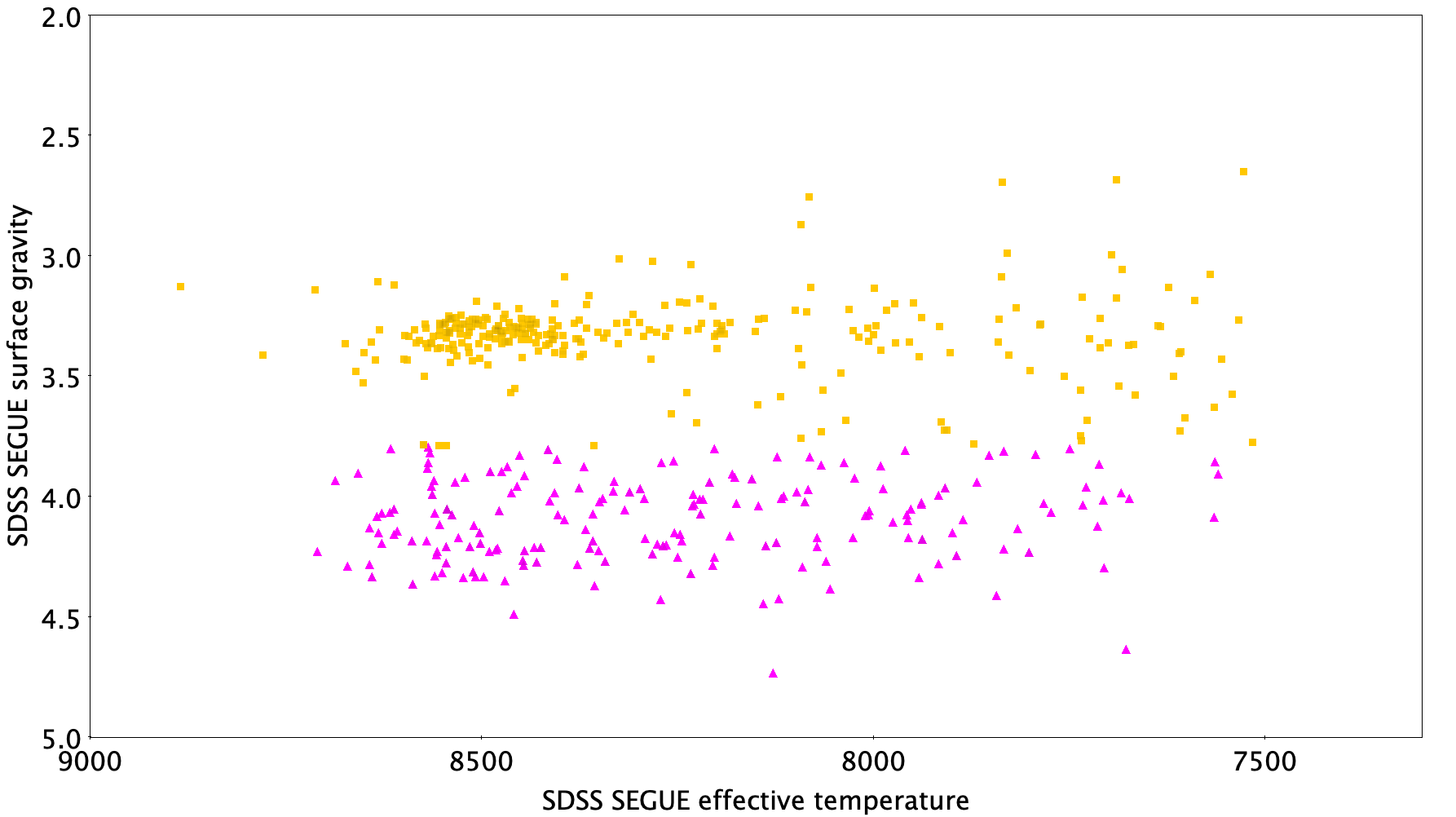}
 \caption{Differentiating between blue horizontal-branch stars and blue stragglers. Upper panel: Gaia EDR3 objects that satisfy the modified (Lindegren et al. 2018) and the final CMD criteria (grey dots) with those with the stringent halo criteria and the crowded region criterion applied (blue circles); SDSS SEGUE objects plotting in the blue horizontal-branch region (yellow squares); and SDSS SEGUE objects plotting in the blue straggler region (magenta triangles) in TEFF-LOGG space. Lower panel: SDSS SEGUE TEFF x LOGG plot with blue horizontal-branch (yellow square) and blue straggler stars (magenta triangles).}
  \label{segue_plots}
  \end{figure}

Cross-matching the parallax candidate objects with the SDSS SEGUE database found 556 stars within the final Gaia CMD criteria (Table 1) and 276 stars within the Galactic halo selection criterion (Table 2). Applying the cut-offs used in \citet{santucci}, we found that 70\% plotted in the blue horizontal-branch region, 10\% plotted in the population I main-sequence region, and 20\% plotted in the blue straggler region (Figure~\ref{segue_plots}).

We found that modifying the final Gaia CMD cut-off from:
\begin{center}
$-0.1 < (G_{BP} - G_{RP}) < 0.53$; \\
$-1 < G_\textrm{abs} < (2.5 - 2.77(G_{BP} - G_{RP}))$
\end{center}
to the more restrictive
\begin{center}
$-0.1 < (G_{BP} - G_{RP}) < 0.53$; \\
$-1 < G_\textrm{abs} < (1.5 - 2.0(G_{BP} - G_{RP}))$
\end{center}
gave a smaller catalogue but with a lower contamination level, with 86\% of objects in the blue horizontal-branch region, 8\% of objects in the population I main-sequence region, and 6\% in the blue-straggler region. This more restrictive selection was not used in the final catalogue, but it demonstrates that a higher level of purity can be achieved at the expense of completeness.

Cross-matching the proper motion candidate objects with the SDSS SEGUE database found 488 stars within the Gaia colour and reduced proper motion criteria (see Section 3.2) and 107 stars within the Galactic halo selection criterion ($b < -50^{\circ}$ or $b > 50^{\circ}$). Applying the same cut-offs, we found that 68\% plotted in the blue horizontal-branch region, 8\% plotted in the population I main-sequence region, and 24\% plotted in the blue-straggler region.

The catalogue of blue horizontal-branch candidate objects is thus estimated to have a purity level of ${\sim}70\%$. The main contaminants are population I main-sequence stars with a contamination level of ${\sim}10\%$, and a blue-straggler contamination of ${\sim}20\%$. Using more restrictive CMD selection criteria can reduce these levels of contamination for the candidate objects selected using parallax measurements, but at the cost of reduced completeness of the catalogue.

It should be noted that cross-matching the catalogues of blue horizontal-branch candidates with the SDSS SEGUE database only finds a relatively small number of matches, which are not evenly distributed over the sky. Thus, the calculated levels of purity and contamination can only be considered to be estimates indicating a general level achieved.

The catalogue does not include those blue horizontal-branch stars that do not conform to the various astrometric and photometric quality criteria applied, or those with close apparent neighbours, or those that exist in the direction of crowded regions. The actual number of such objects has not been estimated.

The distance distribution of parallax selection objects peaks at around $\sim$5.5 kpc and goes up to $\sim$10. kpc. The proportion of parallax candidates that are found at $|b|>50^\circ$ is 22\%. This apparent overpopulation at higher Galactic latitudes (only $\sim$8\% of a sphere's volume is at $|b|>50^\circ$) indicates that the crowded region and apparent neighbour filters, which dominate at lower Galactic latitudes, have a beneficial effect on the catalogue's purity at the expense of the catalogue's completeness. It also implies that neither the catalogue's completeness nor its purity is uniform in all directions.

There are 30,343 of the 336,911 final Gaia EDR3 CMD criteria objects that conform to the $V_T > 145$~km s$^{-1}$ criterion. Applying the crowded region criterion and the near apparent neighbour criterion to these 30,343 objects reduces the number to 16,794. This shows that these criteria have removed roughly half of the potential candidates and will thus have a significant effect on the completeness of the catalogue.

A further adverse effect on the completeness is due to the high line-of-sight radial velocity of some halo objects, as noted in Section 3.2. Using the halo stellar velocity distribution functions given in Section
3.1.3 in \citet{anguiano}, we estimate that $\sim$5\% of halo objects will not pass the $V_T > 145$~km s$^{-1}$ criterion and will, thus, be missing in the catalogue. The number of thin-disk stars with an apparent tangential velocity greater than $145$~km s$^{-1}$ is considered to be negligible.
When we make the assumption all blue horizontal-branch stars have an absolute magnitude of 1, then we can calculate an approximate distance to the reduced proper motion selection objects using:
\begin{center}
$d \approx 10^{(0.2 . (\verb!phot_g_mean_mag! - 1 + 5))}$.
\end{center}
Using this equation, we find that the distance distribution of these objects peaks at $\sim$12 kpc with 50\% of the objects closer than $\sim$22 kpc but with some candidates found out to $\sim$60 kpc.

\section{Catalogue of blue horizontal-branch objects}

\begin{table*}[h!]
\centering
\begin{tabular}{ c c c }
\hline\hline
Table Entry & Units & Description \\ [0.5ex]
\hline
$\verb!source_id!$ & - & Gaia EDR3 unique source identifier \\
$\verb!RA2000!$ & degrees & FK5 J2000.0 right ascension \\
$\verb!DEC2000!$ & degrees & FK5 J2000.0 declination \\
$\verb!parallax!$ & mas & Parallax \\
$\verb!parallax_error!$ & mas & Parallax error \\
$\verb!phot_g_mean_mag!$ & mag & G-band mean apparent magnitude \\
$\verb!bp_rp!$ & mag & BP-RP colour \\
$\verb!abs_g_mag!$ & mag & G-band mean absolute magnitude \\
$\verb!tangential_velocity!$ & km~s$^{-1}$ & Tangential velocity \\
$\verb!pmra!$ & mas~yr$^{-1}$ & Proper motion in the right ascension direction \\
$\verb!pmra_error!$ & mas~yr$^{-1}$ & Standard error of proper motion in the right ascension direction \\
$\verb!pmdec!$ & mas~yr$^{-1}$ & Proper motion in the declination direction \\
$\verb!pmdec_error!$ & mas~yr$^{-1}$ & Standard error of proper motion in the declination direction \\
$\verb!total_proper_motion!$ & mas~yr$^{-1}$ & Total proper motion ($\mu$) \\
$\verb!total_proper_motion_over_error!$ & mas~yr$^{-1}$ & Standard error of total proper motion ($\sigma_{\mu}$) \\
$\verb!reduced_proper_motion!$ & mas~yr$^{-1}$ & Reduced proper motion $(H)$ \\
$\verb!abs_mag_flag!$ & - & BHB star identified using parallax \\
$\verb!reduced_pm_flag!$ & - & BHB star identified using proper motions \\
\hline\hline
\end{tabular}
\caption{Table showing the structure of the online version of the blue horizontal-branch candidate star catalogue}
\label{table:3}
\end{table*}

The ADQL query for the parallax candidate dataset given in Appendix A returns 30,343 Gaia EDR3 sources, of which 21,751 have no apparent neighbours within 5 arcsec in the Gaia database(Figure~\ref{final_parallax_cmd_dist}). The dataset of parallax candidates did not have a cut-off applied for Galactic latitude, but the criterion that there should be no apparent neighbour within 5 arcsec does, however, remove the many candidate objects in the Galactic plane. This means that our selection places a greater emphasis on having an uncontaminated catalogue rather than on a complete full-sky catalogue. The criterion that the parallax error should be $<$~20$\%$ results in a catalogue of objects that reside within a limited distance from Earth. Using the distance calculated from the parallax, we found that the parallax candidates are found at a minimum distance of 200pc, a mean distance of 4,657pc, and a maximum of 11,888pc (Figure~\ref{Dist_Dist_dr2_edr3} upper). The Gaia EDR3 apparent G magnitude of the parallax selection objects ranged from 6.0 mag to 16.1 mag with a mean value of 14.2 (Figure~\ref{AppMag_Dist_dr2_edr3}, upper panel).

  \begin{figure}
  \centering
  \includegraphics[width=\hsize]{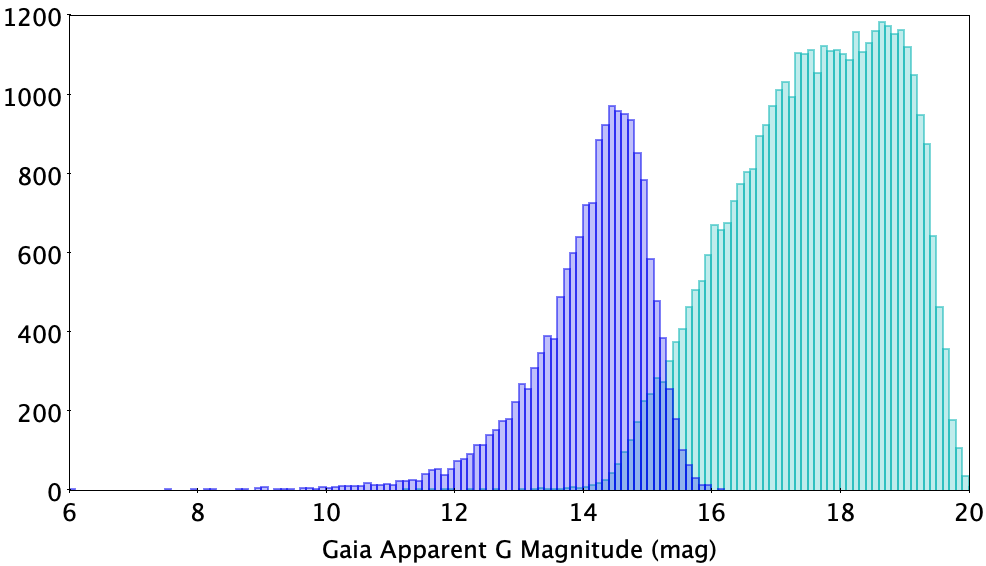}
 \caption{Distribution of Gaia EDR3 apparent G magnitudes of blue horizontal-branch candidate objects selected using parallax (blue) and using proper motions (turquoise).}
  \label{selection_4_rest_dist}
  \end{figure}

\begin{figure}
  \centering
  \includegraphics[width=\hsize]{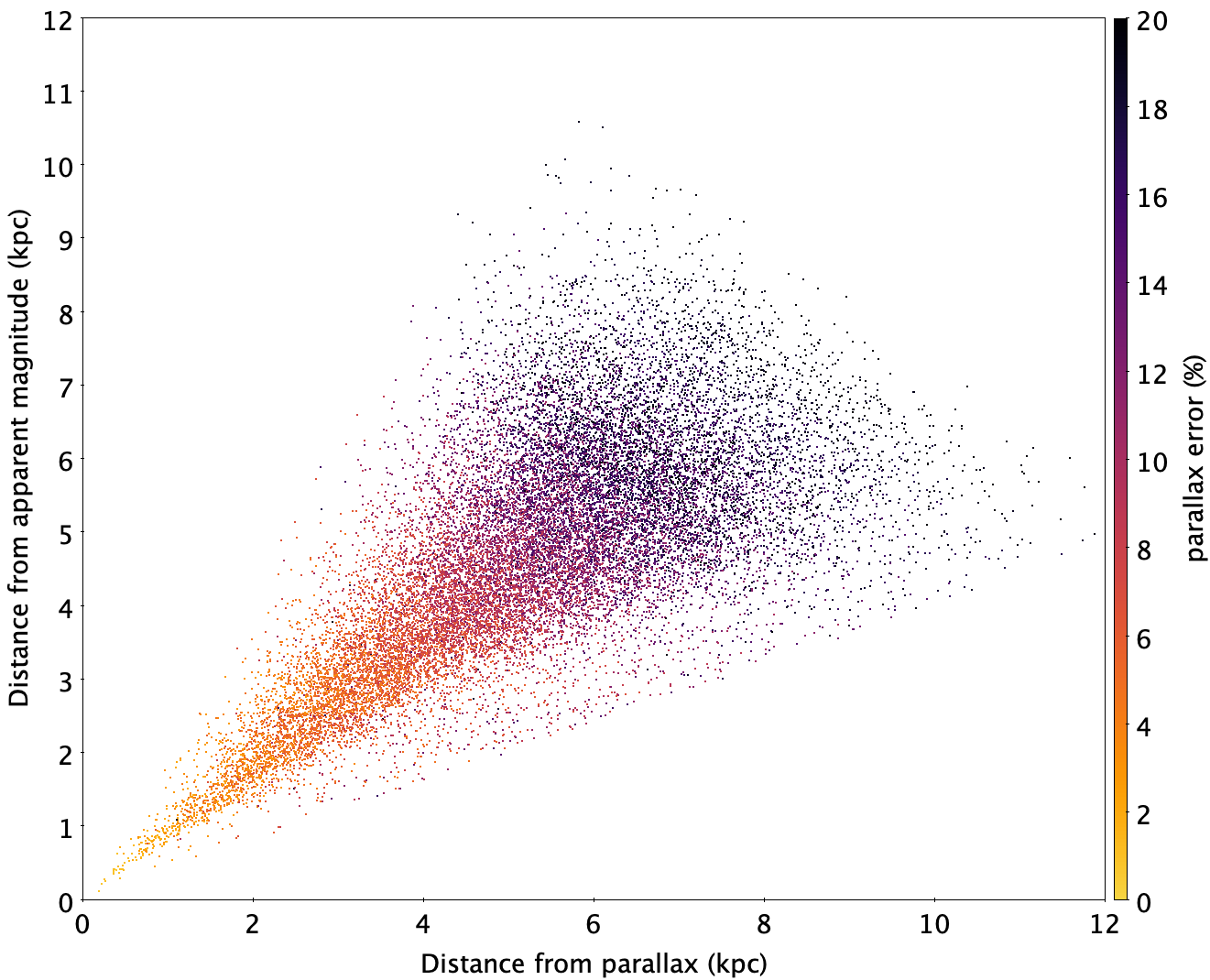}
 \caption{Parallax candidates plotted comparing the distance calculated from parallax to the theoretical distance assuming that all blue horizontal-branch objects have the same absolute magnitude (1 mag). The parallax error is plotted on the colour axis.}
  \label{distance_compare}
  \end{figure}

\begin{figure}
  \centering
  \includegraphics[width=\hsize]{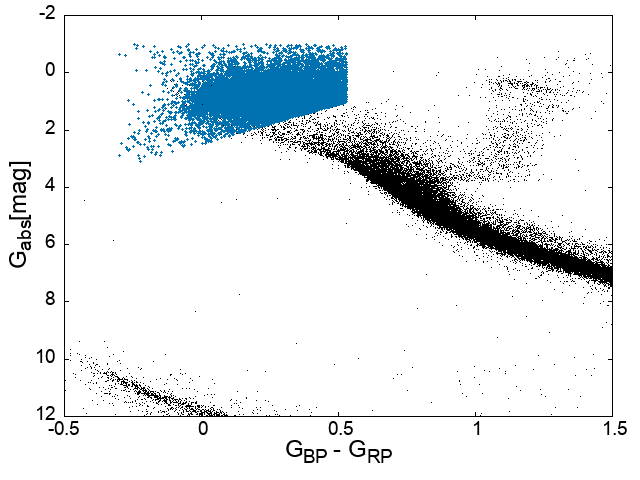}
  \includegraphics[width=\hsize]{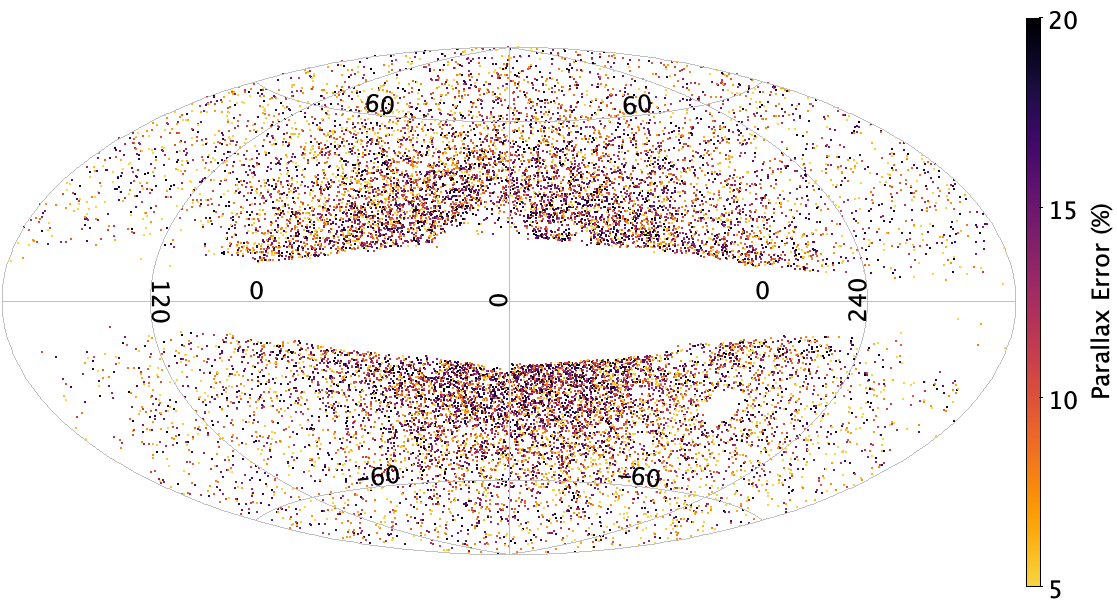}
 \caption{The catalogue of parallax selection objects. Upper panel: Gaia DR2 colour-magnitude diagram showing the blue horizontal-branch parallax candidates (blue) overlain on a \citet{lindegren} 'clean subset selection C' (black). Lower panel: Sky distribution of blue horizontal-branch parallax candidates.}
  \label{final_parallax_cmd_dist}
  \end{figure}

We compared the distance calculated from parallax to the theoretical distance calculated from the apparent magnitude while assuming that blue horizontal-branch stars have a nearly constant absolute magnitude \citep{sirko,deason}. For this purpose, we assumed a constant absolute Gaia EDR3 G magnitude of 1 mag (taken from Figure~\ref{hr_parallax}, upper panel). This approximation is not a state-of-the-art approach, but it demonstrates the distance ranges in question and shows that the increasing scatter with increasing parallax error is as expected (Figure~\ref{distance_compare})

\begin{figure}
  \centering
  \includegraphics[width=\hsize]{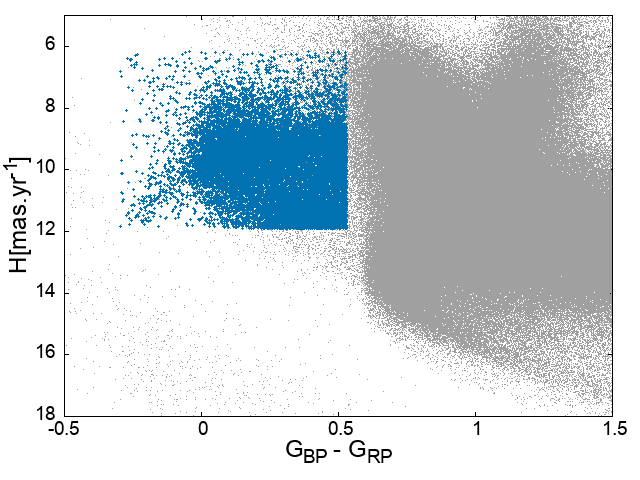}
  \includegraphics[width=\hsize]{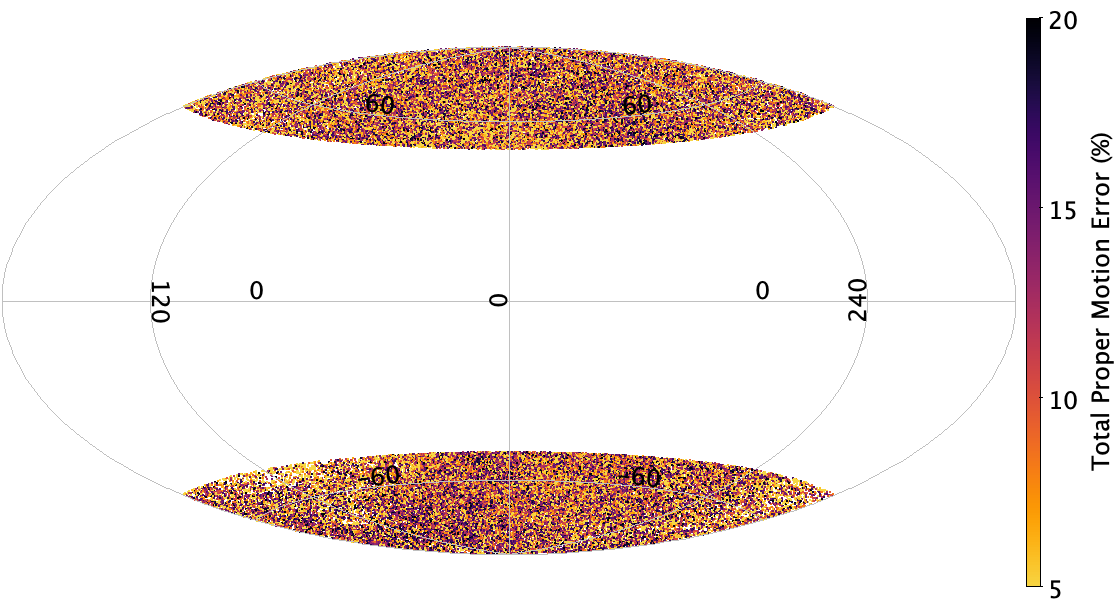}
 \caption{The catalogue of proper motion selection objects. Upper panel: Gaia DR2 colour-reduced proper motion diagram showing the blue horizontal-branch proper motion candidates (blue) overlain on all Gaia DR2 objects with total proper motion errors of <1\% (gray). Lower panel: Sky distribution of blue horizontal-branch proper motion candidates.}
  \label{final_redpm_crpm_dist}
  \end{figure}

The ADQL query for the proper motion data set given in Appendix A returns 42,091 Gaia EDR3 sources, of which 40,538 have no apparent neighbours within 5 arcsec in the Gaia database (Figure~\ref{final_redpm_crpm_dist}). We applied a Galactic latitude cut-off to the proper motion candidates. This was done to ensure that contamination was minimised while consciously accepting that this will not be a complete catalogue. The distribution of Gaia EDR3 apparent G magnitudes ranges from a minimum of 11.3 mag to a maximum of 20.1 mag and a mean of 17.6 mag (Figure~\ref{selection_4_rest_dist}).

\section{Summary and Conclusions}
In this work, we compiled a catalogue comprising of two subsets:
  \begin{enumerate}
 \item 16,794 blue horizontal-branch parallax candidate stars (with parallax error $<$~20$\%$);
 \item 40,583 blue horizontal-branch proper motion candidate stars (with parallax error $\geq$~20$\%$).
  \end{enumerate}
The purity of the catalogue is estimated to be ${\sim}70\%$ with a population I main-sequence contamination level of less than ${\sim}10\%$ seen in both subsets of the final catalogue and a blue straggler contamination levels of ${\sim}20\%$ in both subsets of the catalogue. These purity and contamination estimates have been made using the SDSS SEGUE results.

We used reference datasets \citep{xue,behr} to define the colour and absolute magnitude region of the Gaia DR2 colour-magnitude diagram where blue horizontal-branch stars would be expected. Additional criteria were used based on \citet{lindegren} to ensure that the objects being considered had data that were fit for the purpose. Modifications and additions that were made to the \citet{lindegren} astrometry and photometry quality criteria were based upon similar publications, providing Gaia DR2 based catalogues of extremely low-mass white dwarfs \citep{pelisolivos}, hot sub-dwarf stars \citep{geier}, and white dwarfs \citep{gentilefusillo}.

The population I main-sequence contamination was reduced by focusing on objects in the Galactic halo. The great age of the Milky Way's halo is such that comparatively young main-sequence stars are not expected to be present in large numbers.

These steps were subsequently repeated using Gaia EDR3, having first confirmed that the criteria selected using Gaia DR2 were applicable to Gaia EDR3. A comparison of the Gaia DR2 and Gaia EDR3 based catalogues showed that a richer catalogue has been made possible with the improved accuracy of the astrometric measurements.

Our catalogue is a first step in creating a more complete, full-sky catalogue of blue horizontal-branch stars once more reliable distance measurements as well as more reliable and unambiguous photometry data become available.
It is known that the Herzsprung-Russel diagram on its own is not sufficient to disentangle the blue horizontal-branch from the main-sequence but this catalogue may provide the selection criteria upon which further surveys can be based. In particular, upcoming large spectroscopic surveys (e.g. 4MOST, WEAVE, DESI) can construct their target lists using catalogues such as ours, which identify objects without relying on existing spectra.

\begin{acknowledgements}
IP was partially funded by the Deutsche Forschungsgemeinschaft under grant GE2506/12-1 and by the UK's Science and Technology Facilities Council (STFC), grant ST/T000406/1.

This work has made use of data from the European Space Agency (ESA) mission Gaia (https://www.cosmos.esa.int/gaia), processed by the Gaia Data Processing and Analysis Consortium (DPAC, https://www.cosmos.esa.int/web/gaia/dpac/consortium). Funding for the DPAC has been provided by national institutions, in particular the institutions participating in the Gaia Multilateral Agreement. This research has also made extensive use of TOPCAT (http://www.starlink.ac.uk/topcat/, Taylor
2005), and of NASA’s Astrophysics Data System (http://adsabs.harvard.edu/).
\end{acknowledgements}

%
%

\begin{onecolumn}
\begin{appendix}

\section{ADQL Queries}

For the parallax candidates (parallax error \(<20\%\) dataset):
 \\
SELECT * \\
FROM gaiaedr3.gaia\_source \\
WHERE parallax\_over\_error > 5 \\
AND parallax > 0 \\
AND phot\_bp\_mean\_flux\_over\_error > 10 \\
AND phot\_rp\_mean\_flux\_over\_error > 10 \\
AND phot\_bp\_rp\_excess\_factor < 1.3+0.06*power(bp\_rp,2) \\
AND phot\_bp\_rp\_excess\_factor > 1.0+0.015*power(bp\_rp,2) \\
AND phot\_g\_mean\_mag+5+5*(log10(parallax/1000))>-1 \\
AND phot\_g\_mean\_mag+5+5*(log10(parallax/1000))<(2.5-(2.77*bp\_rp)) \\
AND bp\_rp>-0.1 \\
AND bp\_rp<0.53 \\
AND (sqrt(power(pmra,2)+power(pmdec,2)))*(4.74/parallax)>145 \\
 \\
For the proper motion candidates (parallax error \(\geq20\%\) dataset):
 \\
SELECT * \\
FROM gaiaedr3.gaia\_source \\
WHERE parallax\_over\_error <= 5 \\
AND phot\_bp\_mean\_flux\_over\_error > 10 \\
AND phot\_rp\_mean\_flux\_over\_error > 10 \\
AND phot\_bp\_rp\_excess\_factor < 1.3+0.06*power(bp\_rp,2) \\
AND phot\_bp\_rp\_excess\_factor > 1.0+0.015*power(bp\_rp,2) \\
AND phot\_g\_mean\_mag+5+5*log10((sqrt((pmra*pmra)+(pmdec*pmdec)))/1000)<11.88 \\
AND phot\_g\_mean\_mag+5+5*log10((sqrt((pmra*pmra)+(pmdec*pmdec)))/1000)>6.80 \\
AND (sqrt((pmra*pmra)+(pmdec*pmdec)))/(sqrt((pmdec\_error*pmdec\_error)+(pmra\_error*pmra\_error)))>5 \\
AND bp\_rp>-0.1 \\
AND bp\_rp<0.53 \\
AND abs(b)>=50 \\

The following ADQL call was used to find the number of objects within Gaia DR2 that are within 5 arcsec (0.001388889 deg) of a coordinate:
 \\
SELECT mine.source\_id, count(*) as within\_5\_arcsec\\
from user\_username.filename as mine\\
join gaiadr2.gaia\_source as gaia\\
on 1 = contains(\\
point('ICRS', mine.ra, mine.dec),\\
circle('ICRS', gaia.ra, gaia.dec, 0.00138889)\\
)\\
group by mine.source\_id\\
 \\
This returns a file containing the Gaia DR2 source\_id and the number of objects to be found within 5 arcsec. If there are no apparent neighbours within 5 arcsec then a value of 1 (the star itself) will be returned.

\end{appendix}

\end{onecolumn}

\end{document}